\documentclass[%
 reprint,
superscriptaddress,
nofootinbib,
 amsmath,amssymb,
 aps,
floatfix,
]{revtex4-1}

\usepackage{amsmath}
\usepackage{amssymb}
\usepackage{graphicx}
\usepackage{enumerate}
\usepackage{siunitx}
\usepackage[pdftex,colorlinks=true,citecolor=navyblue,filecolor=navyblue,linkcolor=navyblue,urlcolor=navyblue]{hyperref}
\usepackage{dcolumn}%
\usepackage{bm}%
\usepackage{color}
\usepackage{lineno}
\usepackage{xspace}

\definecolor{navyblue}{rgb}{0.0, 0.0, 0.5}

\graphicspath{ {./fig/} }

\newcommand{\kevnr}{\ensuremath{\textrm{keV}_\textrm{nr}}\xspace} %
\newcommand{\qy}{$Q_{y}$\xspace} %
\newcommand{\ly}{$L_{y}$\xspace} %

\newcommand{\Nee}{N$_{ee}$\xspace}

\newcommand{\cns}{CE$\nu$NS\xspace}

\newcommand{\ty}{t$\times$yr\xspace}

\begin{document}

\title{Nuclear recoil response of liquid xenon and its impact on solar 8B neutrino and dark matter searches}


\author{X.~Xiang} \email[Corresponding author: ] {xxiang@bnl.gov}
\affiliation{Brown University, Department of Physics, Providence, RI 02912-9037, USA}
\affiliation{Brookhaven National Laboratory (BNL), Upton, NY 11973-5000, USA}

\author{R.J.~Gaitskell}
\affiliation{Brown University, Department of Physics, Providence, RI 02912-9037, USA}

\author{R.~Liu}
\affiliation{Brown University, Department of Physics, Providence, RI 02912-9037, USA}

\author{J.~Bang}
\affiliation{Brown University, Department of Physics, Providence, RI 02912-9037, USA}

\author{J.~Xu}
\affiliation{Lawrence Livermore National Laboratory (LLNL), Livermore, CA 94550-9698, USA}

\author{W.H.~Lippincott}
\affiliation{University of California, Santa Barbara, Department of Physics, Santa Barbara, CA 93106-9530, USA}

\author{J.~Aalbers}
\affiliation{SLAC National Accelerator Laboratory, Menlo Park, CA 94025-7015, USA}

\author{J.E.Y.~Dobson}
\affiliation{University College London (UCL), Department of Physics and Astronomy, London WC1E 6BT, UK}

\author{M.~Szydagis}
\affiliation{University at Albany (SUNY), Department of Physics, Albany, NY 12222-1000, USA}

\author{G.R.C.~Rischbieter}
\affiliation{University at Albany (SUNY), Department of Physics, Albany, NY 12222-1000, USA}

\author{N.~Parveen}
\affiliation{University at Albany (SUNY), Department of Physics, Albany, NY 12222-1000, USA}

\author{D.Q.~Huang}
\affiliation{Brown University, Department of Physics, Providence, RI 02912-9037, USA}
\affiliation{University of Michigan, Randall Laboratory of Physics, Ann Arbor, MI 48109-1040, USA}

\author{I.~Olcina}
\affiliation{University of California, Berkeley, Department of Physics, Berkeley, CA 94720-7300, USA}
\affiliation{Lawrence Berkeley National Laboratory (LBNL), Berkeley, CA 94720-8099, USA}

\author{R.J.~James}
\affiliation{University College London (UCL), Department of Physics and Astronomy, London WC1E 6BT, UK}

\author{J.A.~Nikoleyczik}
\affiliation{University of Wisconsin-Madison, Department of Physics, Madison, WI 53706-1390, USA}

\date{\today}


\begin{abstract}
    \noindent
     Knowledge of the ionization and scintillation responses of liquid xenon (LXe) to nuclear recoils is crucial for LXe-based dark matter experiments. Current calibrations carry large uncertainties in the low-energy region below $\sim3$ \kevnr where signals from dark matter particles of $<$10 GeV/c$^2$ masses are expected. The coherent elastic neutrino-nucleus scattering (\cns) by solar $^8$B neutrinos also results in a continuum of nuclear recoil events below {3.0} \kevnr (99\% of events), which further complicates low-mass dark matter searches in LXe experiments. In this paper, we describe a method to quantify the uncertainties of low-energy LXe responses using published calibration data, followed by case studies to evaluate the impact of yield uncertainties on ${^8}$B searches and low-mass dark matter sensitivity in a typical ton-scale LXe experiment. We conclude that naively omitting yield uncertainties leads to overly optimistic limits by factor $\sim2$ for a 6 GeV WIMP mass. Future nuclear recoil light yield calibrations could allow experiments to recover this sensitivity and also improve the accuracy of solar ${^8}$B flux measurements.
\end{abstract}


\maketitle

\newpage

\section{Introduction} \label{sec:intro}


Liquid xenon (LXe) dark matter experiments have greatly improved their sensitivities to rare nuclear recoil (NR) signals in the last decade. As of 2022, there are three multiton-scale LXe experiments in operation and searching for anticipated NR signals from Weakly Interacting Massive Particles (WIMPs) --- LUX-ZEPLIN (LZ)~\cite{LZ:2022ufs}, XENON-nT~\cite{XENON:2020kmp}, and PandaX-4T~\cite{PandaX-4T:2021bab}. Recently the XENON, LZ, and DARWIN collaborations have joined force to study the possibility of building the next-generation LXe observatory for dark matter and neutrino physics~\cite{Aalbers:2022dzr}.

The centerpiece of a typical LXe dark matter detector is a dual-phase xenon time projection chamber (TPC), where particle interactions in the liquid produce prompt scintillation light (S1) at $\sim$$\SI{175}{nm}$ and ionization electrons. The ionization electrons can either recombine with ions to produce additional scintillation light~\cite{Lenardo:2015}, or become liberated and drift away from the interaction site under the influence of an external electric field. Once the electrons are extracted into a gas xenon (GXe) region, they can produce secondary scintillation (S2) via electroluminescence. Combining S1 and S2 signals, LXe experiments have achieved low energy thresholds, accurate position reconstruction in 3D, and strong  discrimination against electron recoil (ER) backgrounds. 

Coherent elastic neutrino-nuclei scattering (\cns) is a neutral current interaction in which a neutrino of any flavor scatters off a nucleus as a whole, producing a nuclear recoil (NR). The process requires the momentum transfer to be significantly smaller than the inverse of the targeted nuclear size, restricting neutrino energies to below a few tens of MeV. The \cns cross-section on a spin-0 nucleus with Z protons and N neutrons at rest without radiative corrections is well described by the Standard Model (neglecting a second order term)~\cite{Freedman:1973yd}:
\begin{equation}
\frac{d\sigma}{dE_r}
	= \frac{G^2_F M_A}{\pi} \left[ 1- \frac{E_r}{E_\nu} - \frac{M_A E_r}{2E_\nu^2} \right] \frac{Q_W^2}{4} F^2(q)
\label{eqn:cevns}
\end{equation}
where $E_r$ is the NR energy, $E_\nu$ is the neutrino energy, $M_A$ is the target mass, $G_F$ is the Fermi coupling constant, $F$ is the nuclear form factor, $q$ is the momentum transfer, and $Q_W = N - (1-4\sin^2 \theta_W) Z$  is the weak nuclear charge with $\theta_W$ being the Weinberg's angle. In 2017, \cns was first experimentally observed using a CsI[Na] crystal at a high energy neutrino beam by the COHERENT collaboration with 6.7-sigma significance~\cite{COHERENT:2017ipa}.

\begin{figure}[!!t]
    \begin{center}
        \includegraphics[width=0.5\textwidth]{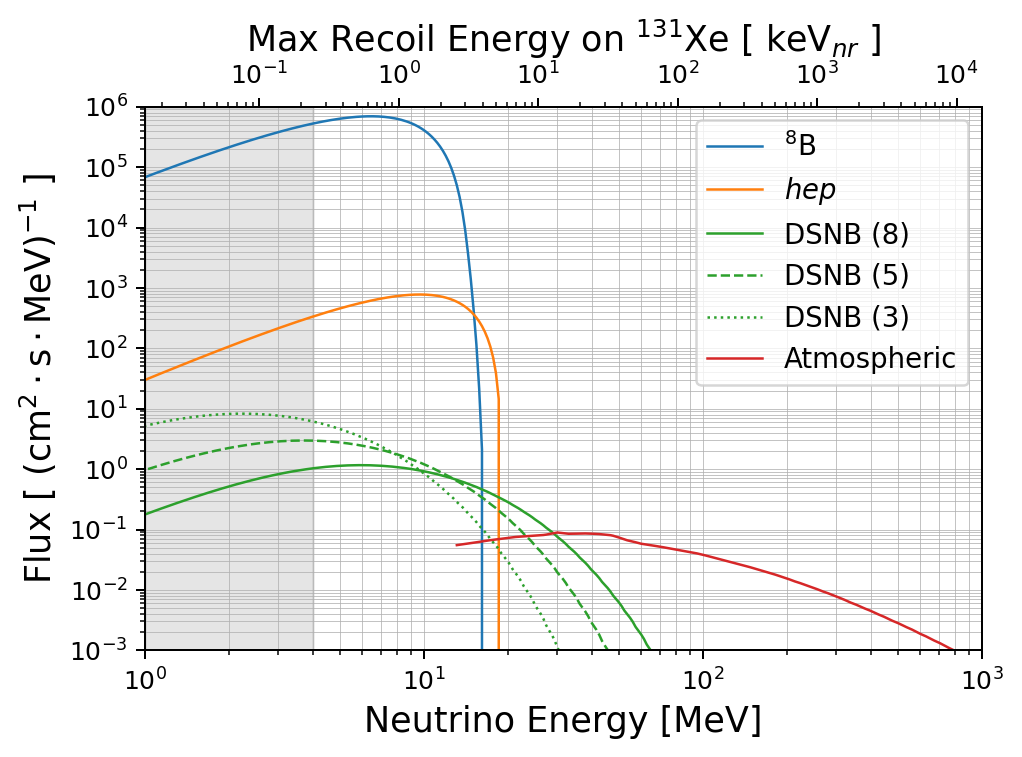}
        \caption{The neutrino flux that can induce observable \cns events in a LXe detector. The grey area corresponds to an energy region that produces 0 observable quanta (photons and electrons) according to the NEST NR yield model, and is beyond the reach of a LXe detector. The Diffused Supernova Neutrino Background (DSNB) spectra are shown at various Fermi-Dirac temperatures in units of MeV \cite{Strigari:2009bq}. Because the sub-GeV atmospheric neutrino remains unexplored by current experiments, the sub-GeV spectrum for dark matter experiments are based on FLUKA simulation. The low-energy  cut off is caused by the lack of low-energy cosmic ray data and interaction uncertainties between cosmic rays and air nuclei as the simulation inputs~\cite{Honda:2011nf,Super-Kamiokande:2015qek}. The top x-axis is the maximum recoil energy from a neutrino back-scattering on a $^{131}$Xe nucleus.}
        \label{fig:nu_flux}
    \end{center}
\end{figure}

Underground LXe TPC experiments are ideal to observe naturally occurring neutrinos via \cns, thanks to the $\sim$$N^2$-enhanced interaction cross-section with a xenon target and their ability to separate \cns from electron recoil (ER) backgrounds. With a sub-\kevnr detection threshold, LXe detectors could observe \cns for neutrino energies down to $\sim$5 MeV, substantially below the neutrino energy used by the COHERENT experiment. Natural sources that can produce \cns\ signals in underground LXe detectors include: solar $^8$B neutrinos, solar $hep$ neutrinos, the diffuse supernova neutrino (DSN) background, sub-GeV atmospheric neutrinos ($atm$), and the neutrinos from core-collapse of supernovae ($E_\nu\sim$ O(10 MeV)). The total flux, in units of cm$^{-2}$ s$^{-1}$, used in this analysis for $^8$B, $hep$, DSN, and $atm.$ neutrinos are: $(5.25 \pm 0.2)\times 10^6$~\cite{SNO:2011hxd}, $7.98(1 \pm 0.3)\times 10^3$, $86 \pm 43$ , and $10.5\pm2.1$~\cite{Baxter:2021pqo}, respectively. Figure~\ref{fig:nu_flux} shows energy spectra for the major sources of neutrinos as well as the maximum predicted Xe recoil energy from each. 

\begin{figure}[!!t]
    \begin{center}
        \includegraphics[width=0.5\textwidth]{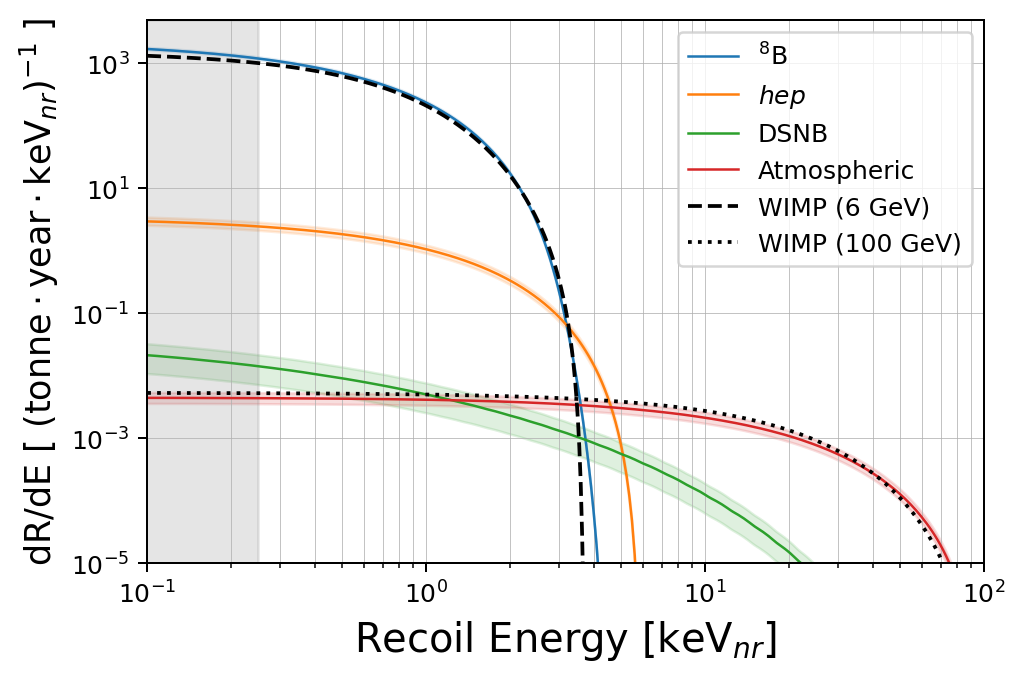}
        \caption{The \cns energy spectrum for all neutrino backgrounds that are relevant to for LXe dark matter experiments. The dashed and dotted spectra are 6 GeV (\SI{4e-45}{cm^2}) and 100 GeV (\SI{3e-49}{cm^2}) WIMPs whose spectrum shapes degenerate with those of $^8$B and atmospheric neutrinos. The grey area corresponds to an energy region that produces 0 observable quanta (photons and electrons) according to the NEST NR yield model. The 1 $\sigma$ uncertainties associated with each neutrino source (transparent color bands) are the neutrino flux uncertainties.}
        \label{fig:drde_cevns}
    \end{center}
\end{figure}

Out of all natural neutrino sources, solar $^8$B neutrinos produce the highest recoil rate in LXe, two orders of magnitude higher than the next most common, $hep$ neutrinos, as shown in Fig. \ref{fig:drde_cevns}. Approximately 750 \cns\ events can be expected from a tonne-year exposure in LXe. However, with a monotonically decreasing energy spectrum, 99\% of the $^8$B \cns\ signals are expected to be below a $\sim3$\kevnr.
In this energy region, the detection efficiency in a LXe TPC is very low. A typical single-scatter interaction in LXe TPCs is required to contain an S1  and S2 pulse separated by the time required to drift the electrons to the gas phase. Usually an identifiable S1 signal requires the simultaneous detection of scintillation light by at least two (2-fold) or three (3-fold) PMT channels. A valid S2 signal typically requires a minimal number of 4 electrons extracted (\Nee) into the gas region~\cite{XENON:2020gfr}.  With these practical energy thresholds taken into consideration, a 5.6 tonne fiducial mass detector like LZ~\cite{LUX-ZEPLIN:2018poe}, which requires at least a 3-fold (2-fold) S1 coincidence and a 5 \Nee S2 threshold, will have the total $^8$B \cns\ rate reduced to $\sim$$1.8$ per \ty ($\sim$$9$ per \ty).

There are both opportunities and challenges that come with the appearance of \cns in LXe detectors. On the one hand, the observation of \cns from solar $^8$B would demonstrate the low-energy nuclear recoil sensitivity of LXe experiments \emph{in situ} and enable a new method to study natural neutrinos at the MeV-scale. For example, \cns provides a unique opportunity to probe non-standard neutrino interactions~\cite{COHERENT:2020iec, Brice:2013fwa}. 
On the other hand, \cns is an irreducible background in WIMP searches, because \cns-induced recoil signatures are indistinguishable from those of WIMPs --- single-scattering NRs uniformly distributed in the active Xe volume with no coincidence signals. In particular, the $^8$B \cns\ spectrum is nearly degenerate with that of a 6 GeV/c$^2$ WIMP (Fig. \ref{fig:drde_cevns}), which weakens an experiment's ability to claim a  discovery of WIMP interactions in this mass region. The degeneracy also amplifies any correlated uncertainty in calculating the WIMP sensitivity for masses around 6 GeV/c$^2$.

In principle, a careful modulation analysis could  resolve the phase difference between WIMP recoils (peak in June 1st~\cite{Freese:2012xd}) and \cns from solar neutrinos (peak at perihelion date around January 3rd), but this method requires plentiful statistics and is beyond the scope of this work. Directional information of the detected nuclear recoils could  be another handle to discriminate between the two signals, but momentum reconstruction has not been demonstrated in any LXe experiment up to date.

As liquid xenon dark matter experiments inevitably march into the solar $^8$B neutrino territory, it is crucial to quantify the common underlying nuclear recoil light yield (\ly) and charge yield (\qy) at energies relevant for observing $^8$B neutrinos. This paper focuses on demonstrating a method to quantify the low-energy \ly and \qy uncertainties, exploring their scientific impacts for a generic LXe dark matter experiment in the presence of $^8$B neutrinos. This paper is organized as follows: Section \ref{sec:model} reviews the published nuclear recoil calibration data on \ly and \qy in LXe, and describes the method we adopt to quantify the low-energy yield uncertainties. Section \ref{sec:phys} demonstrates the impact of the yield uncertainties on physics searches of $^8$B neutrino and low-mass dark matter using a hypothetical LXe detector. Unless stated otherwise, the simulation work presented in this paper assumes a detector with the same performance as that predicted for LZ described in~\cite{LUX-ZEPLIN:2018poe}. Section \ref{sec:discuss} discusses possible strategies to mitigate this source of uncertainty. Finally, Section \ref{sec:summary} summarizes the main results of this paper. 

\section{Modeling Low-energy Yields} \label{sec:model}

The scintillation and ionization mechanisms of LXe are governed by complex micro-physics that cannot be accurately derived from first principles only. The Lindhard theory~\cite{osti_4153115} has been used to estimate the magnitude and fluctuation of LXe's response to NRs. While this theory works well for recoil energies above $\mathcal{O}$(10 \kevnr), deviation from the theory has been observed in a number of measurements in various target materials at low recoil energies. In this work we rely on existing nuclear recoil calibration data in the low energy region and develop a procedure to quantify the uncertainties of the yields.

\subsection{Selection of published data}
\label{sec:model:data}

\begin{figure}[!!t]
    \begin{center}
        \includegraphics[width=0.45\textwidth]{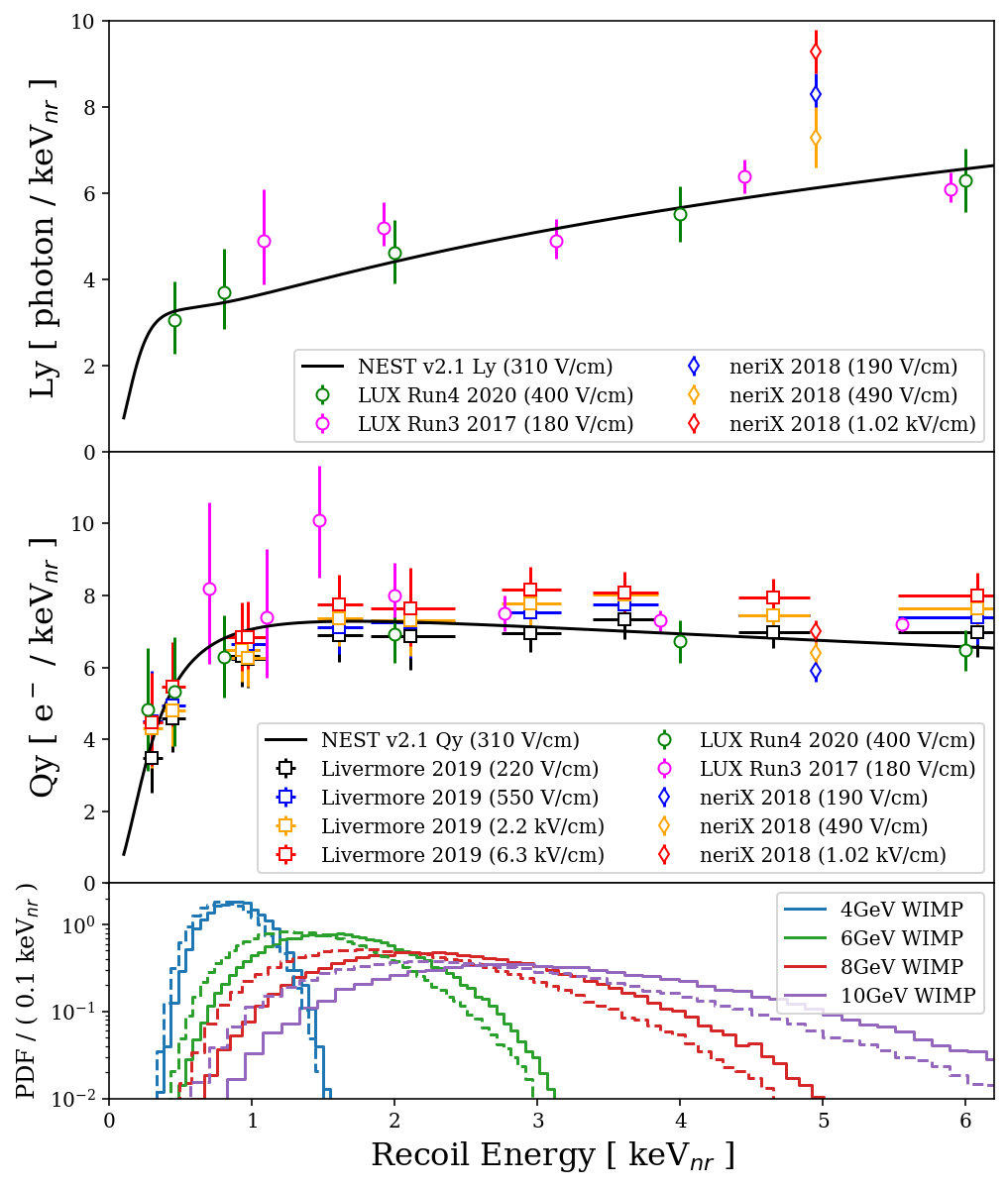}
        \caption{A summary of all NR \ly and \qy measurements at various electric fields published in the past decade. The default NEST v2.1 curves come from a global fit taking into account all available data extending up to 100 \kevnr. Bottom: the detected energy spectra for various low-mass WIMPs computed using NEST v2.1 yield and the detector parameters in~\cite{LUX-ZEPLIN:2018poe}. The solid spectra are for a detector threshold with 3-fold S1 coincidence and $\geq5$ electron S2 requirement. The dashed spectra are for a 2-fold 5 electron detector threshold.}
        \label{fig:yield_data}
    \end{center}
\end{figure}

The Noble Element Simulation Technique (NEST) is a widely used package to model the \ly and \qy responses of xenon to energy depositions within the noble liquid community~\cite{Szydagis:2021}. Multiple experiments have reported \ly and \qy calibration results in the energy region of interest for this analysis (below 3.7 \kevnr)~\cite{Aprile:2018jvg, Huang:2020, LUX:2022qxb, LUX:2016ezw, Lenardo:2019fcn}, which are shown alongside the default NEST yield model v2.1 in Fig \ref{fig:yield_data}. These \ly and \qy results are obtained with neutrons elastically scattering off xenon nuclei. Because current LXe dark matter detectors all operate at drift fields in the range of 100 to 500 V/cm, we leave out the ZEPLIN-III results which were measured at \SI{3.4}{kV/cm}. 
The LUX Run3 \qy data is included in this analysis, but it is found to make a minimal impact on the final results due to its relatively large uncertainty values.

Among the remaining calibration data, the \qy values measured at the Lawrence Livermore National Laboratory (LLNL) using a drift field of \SI{220}{V/cm}~\cite{Lenardo:2019fcn} and those measured in LUX Run4 using a field of \SI{400}{V/cm}~\cite{Huang:2020, LUX:2022qxb} both reached the sub-keV energy scale (down to $\sim$0.3 \kevnr). These two results are consistent with each other with the LLNL data reporting smaller systematic uncertainty values. The data points shown in Fig.~\ref{fig:yield_data} have vertical error bars calculated as the sum of the statistical and the systematical uncertainties in quadrature. Since the \qy data trend exhibits a steep slope below 1 \kevnr, an additional \qy uncertainty is added to the lowest points at $\sim$$0.3$ \kevnr and $\sim$$0.5$ \kevnr to account for recoil energy uncertainties ($\delta E_r$): $\delta Qy = s \delta E_r$, where s = 0.6 electrons/keV$_\mathrm{nr}^2$ is an estimate of the slope.  

For \ly, only the two calibration data sets from LUX are selected and their uncertainties are treated as independent. 
The LLNL experiment did not use reflectors to enhance light collection and thus did not report a simultaneous \ly measurement.
Similar to the treatment of \qy, the \ly error bars in Fig. \ref{fig:yield_data} include statistical and systematical uncertainties summed in quadrature. Because the \ly values are relatively insensitive to NR energy, we do not include energy uncertainties in the \ly error bars.

The response of LXe to a nuclear recoil depends on the electric field strength applied at the interaction site. The original data in Fig.~\ref{fig:yield_data} were collected from experiments under different drift fields. These data sets are corrected to the same field using NEST. For the \ly fitting analysis, a downward scaling is applied to the LUX Run3 \ly values to align them with the LUX Run 4 \ly values at 400 V/cm. The scaling is relatively small, with a maximum shift of 7\% occurring at 1 \kevnr. Similarly, we applied a constant downward scaling factor of 0.969 to the LUX Run4 \qy values to align them with Livermore's \qy value taken at 220 V/cm. This data scaling approach is equivalent to adjusting the drift field parameter in the NEST model. 



\subsection{NEST-based yield model} \label{sec:model:nest}

In the NEST model, the \ly and \qy are defined as the average number of scintillation photons ($N_{ph}$) and the average number of ionization electrons ($N_e$) created per keV of deposited energy. Due to a non-linear quenching effect, the functional forms of the \ly(E) and \qy(E) functions in NEST are empirically driven, where parameters in the yield models are obtained through a simultaneous fit to multiple calibration data sets across a wide energy range. 

Because the measured NR yields below $\sim$\SI{3}{\kevnr} carry much larger uncertainties than those at higher energies, the NEST global fit over a wide energy range is less constrained by the low energy data points and is thus not ideal for this analysis. In addition, there are 11 free parameters in the NEST model, which pose a technical challenge for the error bands of the yields to be properly evaluated. Incorporating such a large number of variables is also computationally expensive in our Profile Likelihood Ratio (PLR) framework, which is commonly used by LXe experiments to calculate experimental sensitivity. 



\label{sec:model:ab}



We make a simplification to the NEST yield model to capture the uncertainties in the low energy region. We denote $Ly^0$ and $Qy^0$ as the default light and charge yield values from NEST version 2.1, and obtain the parameterized $Ly'$ and $Qy'$ by coupling $Ly^0$ and $Qy^0$ to linear terms:

\begin{equation}
    \begin{split}
    Ly' &= Ly^0 (a^{Ly}+b^{Ly}(E_r-E^{Ly}_0)) \\
    Qy' &= Qy^{0} (a^{Qy}+b^{Qy}(E_r-E^{Qy}_0))
    \end{split}
    \label{eqn:ab_model}
\end{equation}
where $E_r$ is the NR recoil energy in the unit of \kevnr and $Ly'$ is the parameterized light yield, and $Qy'$ is the parameterized charge yield as a function of recoil energy. By construction, each yield curve has two independent free parameters $(a, b)$. The $E_0$ parameter serves as a constant offset for the ease of sampling $(a,b)$ points in later steps. We choose $E^{Ly}_0=2$ \kevnr and  $E^{Qy}_0=1$ \kevnr, but the exact values of $E_0$ do not affect the yield values because the parameterization is linear. 
Henceforth, Eqn. \ref{eqn:ab_model} is referred to as the ``$(a, b)$ model."

The nuisance parameters $(a, b)$ allow us to explore different variations of NEST yields; parameter $a$ scales the yields up and down while parameter $b$ introduces an additional degree of freedom by modifying the slopes. This model reduces to the default NEST model when $(a,b)=(1,0)$. We comment that because the NEST model is a global fit to all data, the optimal $(a, b)$ obtained using the selected low-energy data do not necessarily equal to $(1,0)$ exactly, but we anticipate them to not deviate strongly from $(1,0)$.


While the $(a,b)$ model allows us to explore the variations of the entire yield curves, we expect the yields to converge to the NEST predictions beyond 3.4 \kevnr because the yields are well-constrained by high energy measurements. To avoid the over-constraint by high energies, we introduce a sigmoid (or Fermi-Dirac) function $F_D$ coupling to the (a, b) terms as the following:
\begin{equation}
    \begin{split}
 		Ly' &= Ly^0  \left[ \left(a^{Ly}+b^{Ly} (E_r - E^{Ly}_0) \right) F_D + (1 - F_D)  \right] \\
		Qy' &= Qy^0  \left[ \left( a^{Qy}+b^{Qy} (E_r - E^{Qy}_0) \right) F_D + (1 - F_D)  \right] 
        \label{eqn:ab_model_full}
	\end{split}
\end{equation}
The sigmoid is a function of recoil energy $E_r$, and it is defined as:
\begin{equation}
 F_D = \frac{1}{1+e^{(E_r - \mu)/\epsilon}}
\label{eqn: fermi}
\end{equation}
where $\mu$ specifies the location of the transition and $\epsilon$ specifies the slope of the transition. When $E_r\ll\mu$, Eqn \ref{eqn:ab_model_full} is reduced to Eqn. \ref{eqn:ab_model}. When $E_r\gg\mu$, it returns to the default NEST yield model. The values of $\mu$, and $\epsilon$ are summarized in Tab. \ref{tab:ab_parameters}. The choice of $\epsilon=1$ allows a smooth transition toward higher energy.


\begin{figure}[!!t]
    \begin{center}
        \includegraphics[width=0.45\textwidth]{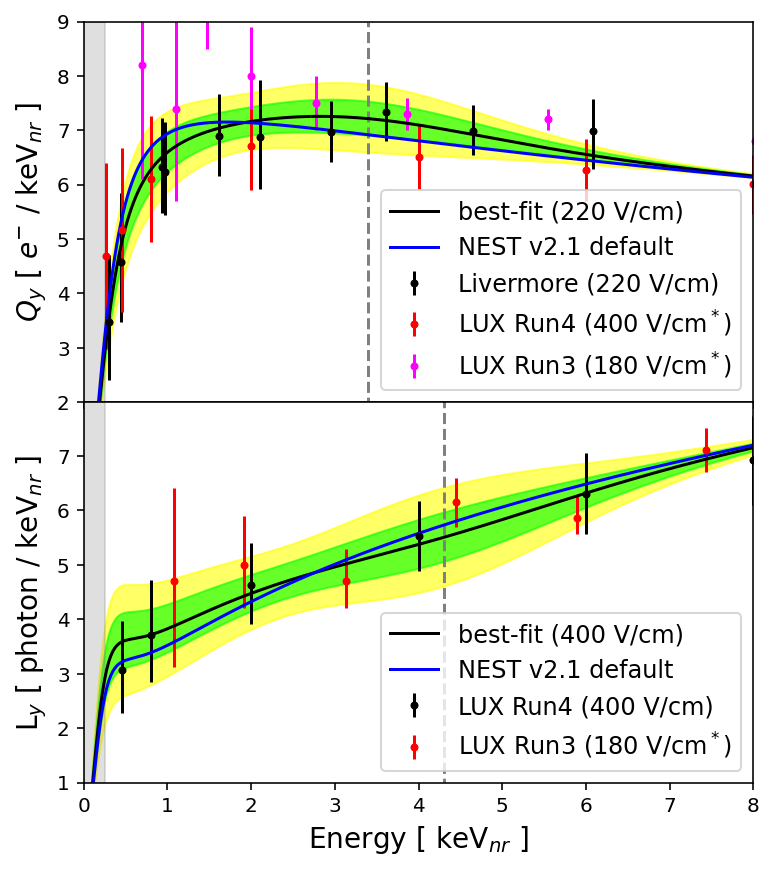}
        \caption{The Brazilian flag uncertainty bands for the yield curves. Top: The bands represent the 1 and 2 sigma uncertainties for \qy propagated from the $(a^\text{Qy},b^\text{Qy})$ model. The Livermore data at 220 V/cm has the smallest uncertainties for \qy.  The * superscript indicates the \qy data points are adjusted to 220 V/cm before the fit. The grey shadow marks the region where a sharp fall-off in NEST is enforced. Bottom: same as the top but for \ly. The LUX Run 4 data at equivalent 400 V/cm has the smallest uncertainties for \ly. The * superscript indicates that the LUX Run 3 \ly data points are adjusted to 400 V/cm before the fit.}
        \label{fig:lyqy_brazilian_flag}
    \end{center}
\end{figure}


To obtain the optimal (a,b) model yields for low energy NRs and the corresponding uncertainties, we use the Pearson $\chi^2$ as the cost function:
\begin{equation}
\chi^2=\sum_i^n \frac{(y_i-f(x_i))^2}{\sigma_i^2}
\label{eqn:cost_function}
\end{equation}
where $x_{i}$ are data points from NR calibrations, $\sigma_i$ are the over all uncertainty values, and $f$ is the parameterized model. After combining systematic and statistical uncertainties in quadrature, we observe rather symmetric error bars (Fig. \ref{fig:yield_data}). The minimum $\chi^2_{min}$ is obtained by performing a simple linear regression fitting of the parameterized model to the best published data. The $(a_{min},b_{min})$ parameters found at the $\chi^2_{min}$ are considered as the median scenario for a yield curve.

The error propagation is handled by the Minuit package~\cite{James:1975dr} to compute \ly and \qy errors across energy: 
\begin{equation}
    \sigma_f = \sqrt{\left(\frac{\partial f}{\partial a}\right)^2 \sigma_a^2 + \left(\frac{\partial f}{\partial b}\right)^2 \sigma_b^2 + 2\rho_{ab} \sigma_a \sigma_b \frac{\partial f}{\partial a}\frac{\partial f}{\partial b}}
    \label{eqn:error_propagation}
\end{equation}
where the values of Hesse errors $\sigma_a$, $\sigma_b$, and correlation $\rho_{ab}$ are summarized in the Tab. \ref{tab:ab_parameters}. The resulting uncertainty bands are shown in Fig \ref{fig:lyqy_brazilian_flag}.

\begin{table}[]
\caption{Summary of parameters used in the $(a,b)$ model.}
\begin{tabular}{c|ccc|ccccc}
\hline \hline
   & \multicolumn{3}{c|}{Constants} & \multicolumn{5}{c}{Fit Results}                            \\
   & $E_0$  & $\epsilon$  & $\mu$  & $a_{min}$ & $b_{min}$ & $\sigma_a$ & $\sigma_b$ & $\rho_{ab}$ \\
   \hline
Qy & 1.0      & 1.0       & 3.4    & 0.92      & 0.05      & 0.060      & 0.052      & -0.585      \\
Ly & 2.0      & 1.0       & 3.4    & 1.04      & -0.05     & 0.086      & 0.072      & -0.354      \\ 
\hline \hline
\end{tabular}
\label{tab:ab_parameters}
\end{table}



\subsection{Simplified yield model}
\label{sec:model:a}


We also consider a simpler version of the $(a,b)$ parameterization with one free parameter in each yield function:
\begin{equation}
    \begin{split}
    Ly' &= Ly^0 + a^{Ly} \\
    Qy' &= Qy^{0} + a^{Qy}
    \end{split}
    \label{eqn:a_model}
\end{equation}

This is similar to the XENON-1T yield model, which uses one free ``interpolation parameter" to uniformly shift the \qy curve and a free coupling parameter to scale the \ly curve~\cite{XENON:2020gfr}. 
We refer to this parametrization as the $a$ model for short, and use the same treatment as the $(a,b)$ model to smoothly transition to the default NEST values at higher energies. The uncertainty bands for this simple model are shown in the Appendix. 

\section{Physics case studies}
\label{sec:phys}


We perform three case studies to evaluate the impacts of the yield uncertainties on physics searches. All three studies assume a hypothetical LXe TPC detector with the same operation condition and background levels as outlined in sensitivity studies of the LUX-ZEPLIN (LZ) experiment~\cite{LUX-ZEPLIN:2018poe}. 
To highlight the effects of the yield uncertainties, we only consider physical backgrounds in this work although instrumental backgrounds including pathological electron emission and accidental backgrounds are reported in majority of currently active xenon TPCs experiments.

\subsection{Detection efficiency} \label{sec:phys:eff}

\begin{figure}[!!t]
    \begin{center}
        \includegraphics[width=0.5\textwidth]{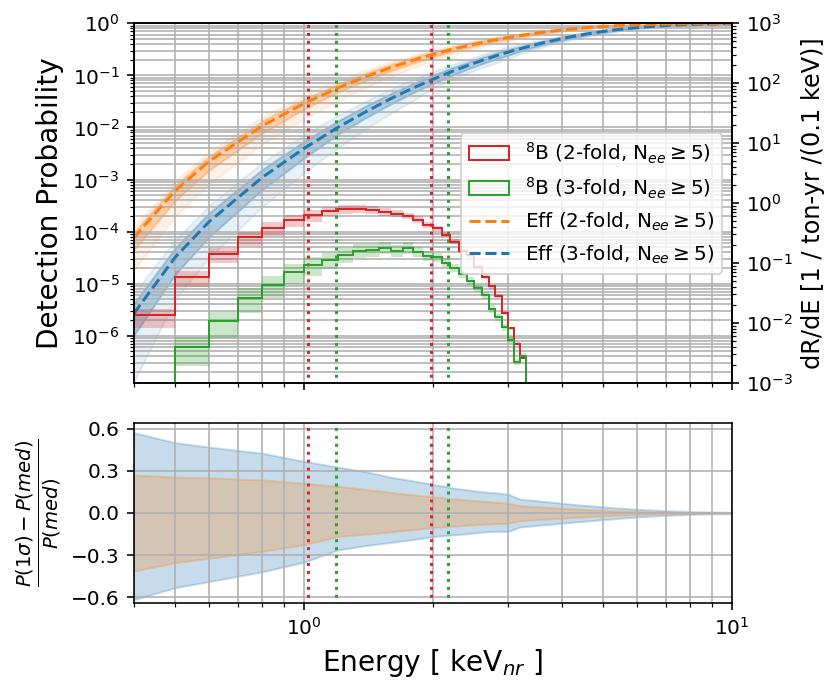}
        \caption{Top: the median NR detection efficiency (dashed) as a function of true recoil energy for two sets of S1 + S2 detector thresholds. The two shaded color bands are the 68\% and 95\% uncertainties from \ly. On the same plot, we show differential rate spectra for solar $^8$B \cns surviving the detection thresholds, whose normalization is indicated by the y-axis on the right. The shaded blue and red bands are derived from the 1$\sigma$ uncertainty in detection efficiencies. The region between two vertical dotted lines indicate where the central 68\% of the events are. Bottom: the fractional uncertainties, defined as the difference from the $\pm 1\sigma$ probability to the median probability divided by the median probability, as a function of energy.}
        \label{fig:nr_threshold}
    \end{center}
\end{figure}

Low-energy NRs generally produce a small number of photons and electrons in the xenon target, which can be challenging to detect in a xenon TPC experiment. An inaccurate quantification of detection efficiency uncertainties would compromise an experimental sensitivity to signals that produce weak NRs. We calculate the signal efficiency at a specific recoil energy as the
fraction of uniformly distributed mono-energetic NR events that survive the detector threshold cuts. 

The yield uncertainties are propagated into the detection efficiency by sampling the yield models  following a $\chi^2$ distribution, as explained below.
First, with the yield model parameterization explained in Sec. \ref{sec:model:ab}, we randomly pick $(a,b)$ values uniformly distributed around $(1,0)$ and calculate the corresponding $\chi^2$ value (Eqn. \ref{eqn:cost_function}). Then we accept the yield model at each $(a,b)$ point with a probability determined by its distance to the minimum, $\delta \chi^2 = \chi^2-\chi^2_{min}$, using the $\chi^2$ PDF with 2 degrees of freedom. 
Next, NEST simulations are performed for each \ly and \qy model and produce an estimate of the detection efficiency at each NR energy, 
based on which the spread of NR detection efficiency is computed. 
The uncertainty bands in Fig. \ref{fig:nr_threshold} represent the central 68\% (heavily shaded) and 95\% (lightly shaded) quantiles of the efficiency curves. 
We stop the calculation at 0.4 \kevnr due to the prohibitive computation time and the impractically low detection efficiency.

Fig. \ref{fig:nr_threshold} also shows the $^8$B energy spectra convolved with the efficiency curves to illustrate the relevant energy region for $^8$B. When the S1 coincidence threshold requirement is relaxed from the 3-fold to 2-fold, Fig. \ref{fig:nr_threshold}-Top shows a drastically improved efficiency that results in an overall increase of $^8$B event rate by a factor of $\sim 5$, while Fig. \ref{fig:nr_threshold}-Bottom shows a significantly reduced relative uncertainty in efficiency. Consequently, lowering the S1 coincidence threshold in a LXe detector is advantageous to probe the low-energy NR phenomenon in the absence of additional background events.

\subsection{Constraint of $^8$B flux}
\label{sec:phys:b8}

A precise measurement of $^8$B \cns\ rate could reveal new physics beyond the Standard Model. This study evaluates an experiment's capability to measure the $^8$B neutrino flux using Monte Carlo-simulated data in the absence of WIMPs. 
The standard \cns interaction with the Helm nuclear form factor was adopted, and seasonal flux modulations are not considered in this analysis.
The detector signals are generated using the default \ly and \qy models in NEST v2.1 but the statistical analysis assume that the yield values are uncertain with spreads estimated in Sec. \ref{sec:model}. 

\begin{figure}[!!t]
    \begin{center}
        \includegraphics[width=0.5\textwidth]{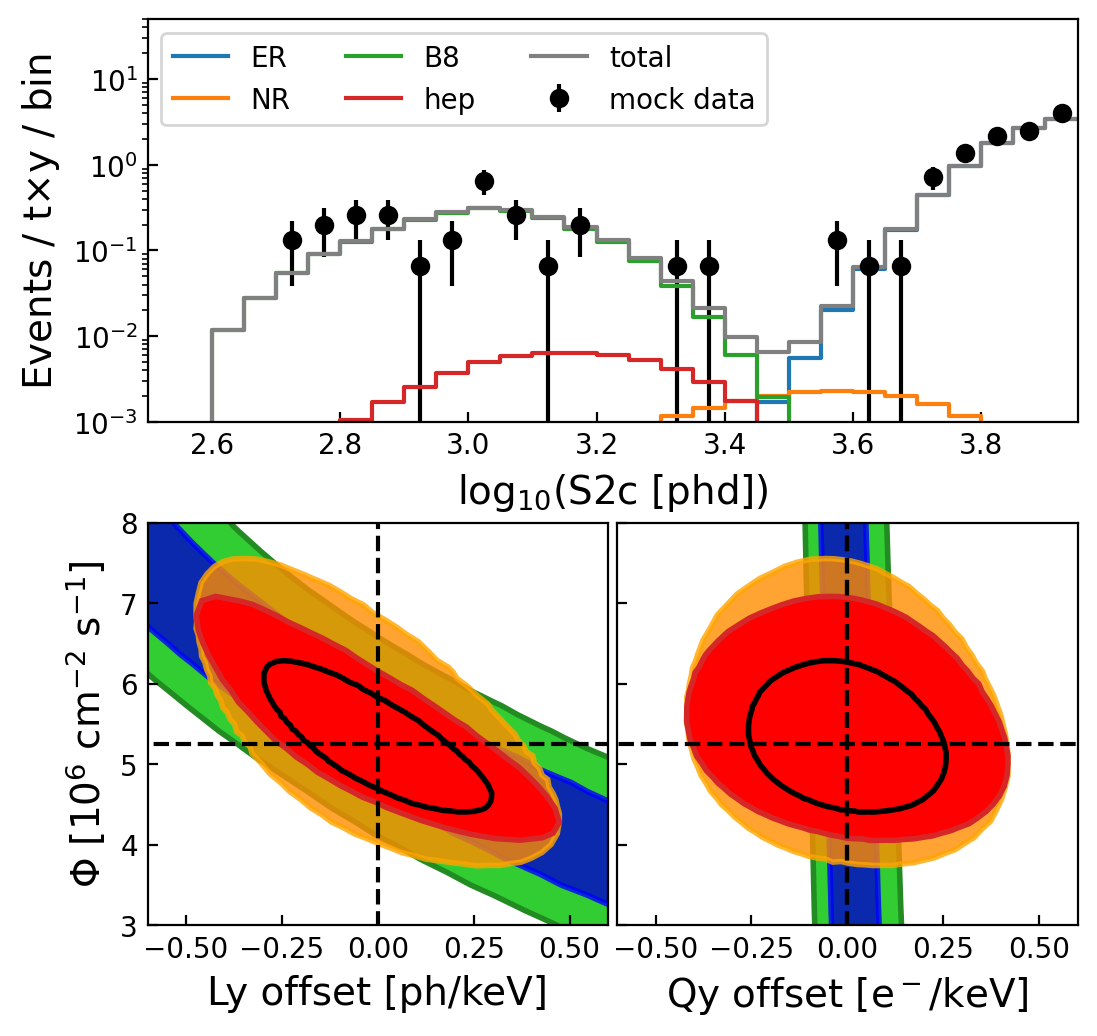}
        \caption{Top: the mock data generated for a 15.3 \ty exposure using the nominal $^8$B flux ($\Phi$ = \SI{5.25e6}{cm^{-2} s^{-1}})~\cite{SNO:2018fch} and the default NEST \ly and \qy curves. The detector thresholds for S1 is 3-fold coincidence and for S2 is 5 \Nee. The error bars are Poissonian. The ER component combines all background labeled as ``ER" in Tab. IV of~\cite{LUX-ZEPLIN:2018poe}. The NR combines all ``NR" labeled background except for solar $hep$ neutrino. The bin width is $0.05 \log(phd)$. Bottom: projections of the 90 \% confidence volumes in $\theta_{Ly}$ and $\Phi$ parameter space. Green (blue) stripe shows the confidence interval for a 15.3 \ty (50 \ty) exposure without any constraints of \ly and \qy. The orange (red) contour shows the interval for a 15.3 \ty (50 \ty) exposure with \ly and \qy constrained to the calibration data using the model in Sec. \ref{sec:model:a}. The solid black line is a 50 \ty exposure where we artificially reduce \ly and \qy uncertainties to half.}
        \label{fig:b8_flux}
    \end{center}
\end{figure}

Fig. \ref{fig:b8_flux} (top) shows the S2 distribution of the simulated data set, along with background PDFs generated using the NEST v2.1 package. The simulation assumes an S1 threshold of 3-fold coincidence and an S2 threshold of 5 extracted electrons. A typical WIMP analysis in a LXe detector is performed in (S1, S2) observable space. 
In this analysis, however, beyond the trigger requirement, including S1 in the analysis does not lead to a significant improvement in the result  due to the small S1 signal amplitudes for both signals and background in this energy region. 
Therefore, this study projects the 2-dimensional data distribution onto the one-dimensional S2 observable space before performing any inference analysis. 

A significant excess of events due to $^8$B signals can be observed clearly over known sources of physical backgrounds, as shown in Fig. \ref{fig:b8_flux} (top). As discussed above (Sec.~\ref{sec:phys}), all LXe TPCs observe some levels of instrumental backgrounds, often called accidental backgrounds, arising from accidental coincidence of spurious S1s (e.g. PMT dark count coincidence) and S2s (e.g. grid electron emission). The characteristics of these backgrounds vary from experiment to experiment, and are still being actively studied~\cite{LUX:2020yym, XENON:2020gfr, Akerib:2021pfd}, making a generic model unreliable. For one example, the XENON1T experiment reported accidental backgrounds at 0.47 events / \ty rate in a 1.3 t fiducial mass but only 0.08 events / \ty in a 0.65 t core mass~\cite{XENON:2018voc}. Here we restrict our study to the known physical backgrounds to highlight the effect of the low-energy yield uncertainties, leaving out the instrumental backgrounds.


To construct confidence intervals, we perform an extended unbinned maximum likelihood fits of the mock data using the Minuit2 package. The model has three parameters of interest: the $^8$B flux $\Phi$, a \ly parameter $\theta_{Ly}$, and a \qy parameter $\theta_{Qy}$, where $\theta_{Ly}$ and $\theta_{Qy}$ provide offsets to shift the default NEST yield curves up and down as prescribed in Sec. \ref{sec:model:a}. Although both $\theta_{Ly}$ and $\theta_{Qy}$ can affect the amplitude and shape of NR PDFs, significant changes in shape only occur under extreme variations of \ly and \qy outside the yield boundaries constrained by published data. Therefore we only allow the amplitude to vary in this work. 

The full log-likelihood function consists of two parts --- an extended event likelihood term ($\mathcal{L}^\text{event}$) and a profile term ($\mathcal{L}^\text{profile}$) to constrain the nuisance parameters ($\boldsymbol{\theta}$) by their associated uncertainties:
\begin{equation}
\log\mathcal{L}^\text{total}(\Phi, \mathbf{\boldsymbol{\theta}}) = \log \mathcal{L}^\text{event} (\Phi, \mathbf{\boldsymbol{\theta}}) + \log  \mathcal{L}^\text{profile}(\boldsymbol{\theta})
\label{eqn:likelihood}
\end{equation}
where $\Phi$ is the $^8$B flux, and the two nuisance parameters are $\theta_{Ly}$ and $\theta_{Qy}$. For a given data set $\mathcal{D}_\text{obs}$, the event likelihood is defined as the following:
\begin{equation}
\begin{split}
\mathcal{L}^\text{event} (\Phi, \mathbf{\boldsymbol{\theta}} | \mathcal{D}_\text{obs}) &= \text{Pois}(n_\text{obs}; \mu_\text{tot}) \\
\times &\prod_{i=1}^{n_\text{obs}} \left[  \mu_s(\Phi, \boldsymbol{\theta})  f_s(\boldsymbol{x}_i) + \sum_{b}\mu_b(\boldsymbol{\theta}) f_b(\boldsymbol{x}_i )\right] 
\end{split}
\label{eqn:event_likelihood}
\end{equation}
where the first term is a Poisson distribution, $n_\text{obs}$ is the number of observed events in this data set, $\mu_\text{tot}$ is the total expected number of events, $\mu_s$ is the expected signal number, $\mu_b$ is the expected background number, and $f_s(\boldsymbol{x}_i)$ and $f_b(\boldsymbol{x}_i)$ are the normalized signal and background PDFs in the observable space $\boldsymbol{x}$ (i.e. logS2). By definition, $\mu_\text{tot} = \mu_s + \sum_b \mu_b$. The signal number $\mu_s$ is a function of $\theta_{Ly}$ and $\theta_{Qy}$, while all expected background number $\mu_b$ are fixed to their respective nominal values. 

The profile term are used to constrain the yield nuisances as the following:
\begin{equation}
\mathcal{L}^\text{profile} (\boldsymbol{\theta}) = \mathcal{G} (\theta_{Ly}; 0, \sigma_{Ly}) \times \mathcal{G} ( \theta_{Qy}; 0, \sigma_{Qy}) 
\label{eqn:b8_profile_term}
\end{equation}
where $\mathcal{G}$ are the Gaussian whose width $\sigma_{Ly}$ ($\sigma_{Qy}$) are derived from fitting external \ly (\qy) calibration data. 

Figure \ref{fig:b8_flux} (bottom) shows the estimated 90\% confidence intervals for the obtained $^8$B flux. The green and blue stripes represent the results when the \ly and \qy constraints are removed; in this case a measurement of the $^8$B flux is impossible due to the strong correlation between the flux and the yield values. When constraints in \ly and \qy are imposed using NR calibration data, the yield-flux degeneracy is broken and measurements of the $^8$B flux can be achieved for projected exposures of 15.3 \ty (orange) and 50 \ty (red). The black contour is for the hypothetical scenario where the \ly and \qy uncertainties~\cite{Lenardo:2019fcn, Huang:2020, LUX:2022qxb} are reduced by half, illustrating how improved calibrations can lead to better confined confidence intervals for the $^8$B flux in an experiment.

\subsection{WIMP sensitivity}

In this section, we evaluate the projected WIMP sensitivity (i.e., the exclusion limit to reject elastic spin-independent WIMP-nucleon scattering at a 90\% confidence level) for the hypothetical LXe experiment in the presence of the $^8$B background with uncertain experimental \ly and \qy yield values.

The sensitivity is calculated using the Profile Likelihood Ratio (PLR) software developed by the LZ collaboration. This analysis follows the convention recommended by~\cite{Baxter:2021pqo} to use a two-sided Frequentist test statistics, a strictly positive signal strength estimator, and avoids the Asymptotic approximation. The WIMP interaction rate is calculated using the Helm nuclear form factor and the standard halo model with the astrophysical parameter values~\cite{Lewin:1995rx,McCabe:2013kea, Smith:2006ym, Schoenrich:2009bx, Bland_Hawthorn_2016} summarized in~\cite{Baxter:2021pqo}.




The event model is constructed from two-dimensional PDFs in (S1, logS2) for each background and signal source.
We use the same background models as explained in~\cite{LUX-ZEPLIN:2018poe} but merge similar background PDFs together to increase the computation speed so we can explore multiple scenarios with different yield assumptions. In particular, all ER background PDFs are combined after weighing each component by its expected event rate; the atmospheric neutrino ($\nu_{atm}$) and diffused supernova neutrino ($\nu_{DSN}$) are absorbed into the environmental neutron background PDF. This simplification is not expected to affect the final outcome because these backgrounds have minimal contamination in the signal region: $<$0.1 events of ER backgrounds may leak into the 90\% event contour of a 10 GeV/$c^2$ WIMPs in 15.3 \ty of exposure; $\nu_{atm}$ and $\nu_{DSN}$, when combined, only contribute 0.024 events/\ty in the energy range from .25 to 6.0 \kevnr. Similar to the previous case study, instrumental backgrounds are not considered here.

The construction of the likelihood function 
follows Eqn. \ref{eqn:likelihood} and Eqn. \ref{eqn:event_likelihood}, except that the flux $\Phi$ is replaced by the WIMP-nucleon cross-section $\sigma_{\chi-N}$, and the background rates $\mu_b(\theta)$ are no longer constants. Since the PDF shapes of low-mass WIMPs ($\leq10$ GeV) and $^8$B backgrounds are insensitive to modest yield variations near the measured values, we define the \ly nuisance parameters as relative scaling factors ($\theta^\text{Ly}_s, \theta^\text{Ly}_b$). These parameters are coupled to the signal and background rates as $\mu_s = \mu_s^0 \theta^\text{Ly}_s$ and $\mu_\text{b} = \mu^0_\text{b} \theta^{\Phi}_\text{b} \theta^\text{Ly}_\text{b}$, where $\theta^{\Phi}_b$ nuisance account for $^8$B flux uncertainty, $\mu_\text{s}^0$ and $\mu^0_\text{b}$ are the nominal WIMP and $^8$B rates in the absence of uncertainty. This treatment was verified to  not affect the outcome while 
simplifying the PLR computation.

A profile term is defined to capture the strong correlation between the WIMP signal and the $^8$B background as a result of low-energy yield uncertainties:
\begin{equation}
\mathcal{L}^\text{profile} (\boldsymbol{\theta}) = \prod_b \mathcal{G}_b (\theta_b ; a_b, \sigma_b)  \times \mathcal{C} (\theta^\text{Ly}_s, \theta^\text{Ly}_\text{8B})
\label{eqn:profile}
\end{equation}
where each $\mathcal{G}_b$ is a Gaussian constraint for a background rate indexed as $b$, and $\mathcal{C}$ is a 2-dimensional analytic constraint function 
describing the correlation between low-mass WIMPs at a specific mass value and $^8$B neutrinos as a result of the \ly variation.
The correlation function is obtained from NEST simulations ahead of the PLR computation 
using a procedure similar to that described in Sec.~\ref{sec:phys:eff}. For each set of $(a,b)$ parameters sampled according to their probability distribution function, we simulate $^8$B and WIMP events in (S1, logS2) space using NEST. Each NEST run returns a $^8$B rate and a WIMP rate, which are recorded in a 2-dimensional histogram. Then we 
obtain a distribution that describes the correlation between the two scaling factors $\theta^\text{Ly}_s$ and $\theta^\text{Ly}_b$ arising from the yield uncertainties. Finally, we fit an empirical function $\mathcal{C}$ to describe the distribution smoothly. The following functional form works well in our case:
\begin{equation}
    \mathcal{C} (\theta^\text{Ly}_s, \theta^\text{Ly}_\text{8B}) = \mathcal{G}(\theta^\text{Ly}_\text{8B}) \mathcal{G}(\theta^\text{Ly}_s | m(\theta^\text{Ly}_\text{8B}), w(\theta^\text{Ly}_\text{8B}))
    \label{eqn:ly_constraint}
\end{equation}
where the second Gaussian is a conditional PDF whose mean $m$ quadratically depends on $\theta^\text{Ly}_\text{8B}$ and width $w$ linearly depends on $\theta^\text{Ly}_\text{8B}$. The fitting is easier done in two steps --- first we fit the x-projection of the histogram with $\mathcal{G}(\theta^\text{Ly}_\text{8B})$, and then we proceed to fit the entire 2D histogram with the 2 parameters fixed. 
The profiling term for \qy follows the same construction. 

\begin{figure}[t]
    \begin{center}
        \includegraphics[width=0.45\textwidth]{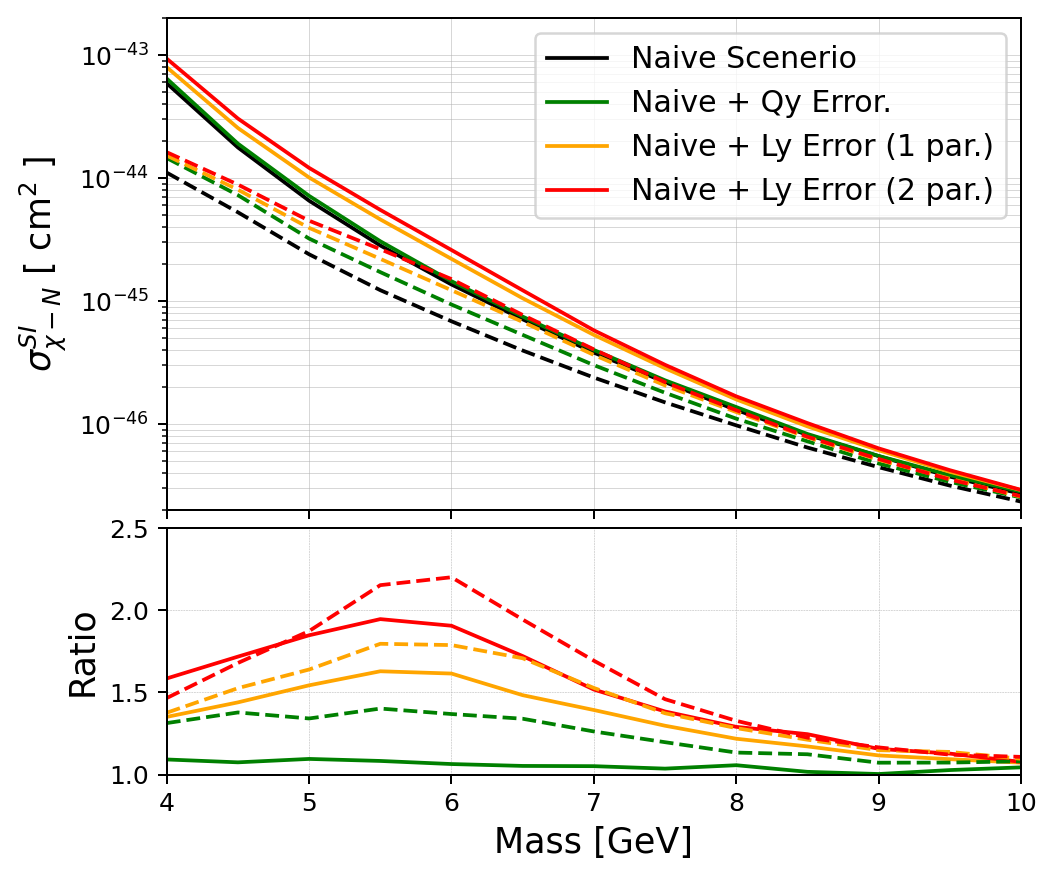}
        \caption{Top: The projected median sensitivity (90\% confidence limit) for a 15.3 \ty of LXe exposure as a function of WIMP mass. The solid curves assume the standard 3-fold 5 \Nee detector threshold for the LZ-like detector, and dashed curves assumes a relaxed 2-fold threshold 5 \Nee detector threshold. The black curves are the baseline scenario where neither \ly nor \qy uncertainty is included. Green curves include \qy uncertainty only, orange curves include \ly uncertainty derived from the single parameter model (Sec. \ref{sec:model:a}), red curves include \ly uncertainty derived from $(a,b)$ models (Sec. \ref{sec:model:ab}). Bottom: the ratio, computed by taking the colored curves from the top plot, divided by the black curves (baseline). }
        \label{fig:limit_vs_mass}
    \end{center}
\end{figure}

The evaluated median WIMP sensitivities for a projected 15.3 \ty exposure are shown in Fig.~\ref{fig:limit_vs_mass}. The black curves represent the ideal scenario where both the \ly and \qy yields are precisely known, and the colored curves include different yield uncertainties,  which produce worsened limits as expected. Three observations can be made from Fig.~\ref{fig:limit_vs_mass}. First, the yield uncertainties impact WIMP masses around 5.5-6 GeV the most due to the degeneracy between $^8$B \cns\ and WIMP spectra for those masses. For WIMP masses around 10 GeV/$c^2$ or higher, the effect of the yield uncertainties diminishes and can be safely neglected, because the WIMP spectra significantly diverges from that of $^8$B. 
Second, the WIMP sensitivity obtained with a 2-fold S1 coincidence threshold is more susceptible to the yield uncertainties than that with a 3-fold threshold, even though the 2-fold threshold generally leads to improved limits. This may be explained as the larger yield uncertainties at lower NR energies. 
Third, the impact of \ly on WIMP sensitivities dominates over that of \qy due to the larger \ly uncertainty in available calibration data, which highlights the need to improve \ly accuracy in future NR calibrations. 

\begin{figure}[!!t]
    \begin{center}
        \includegraphics[width=0.45\textwidth]{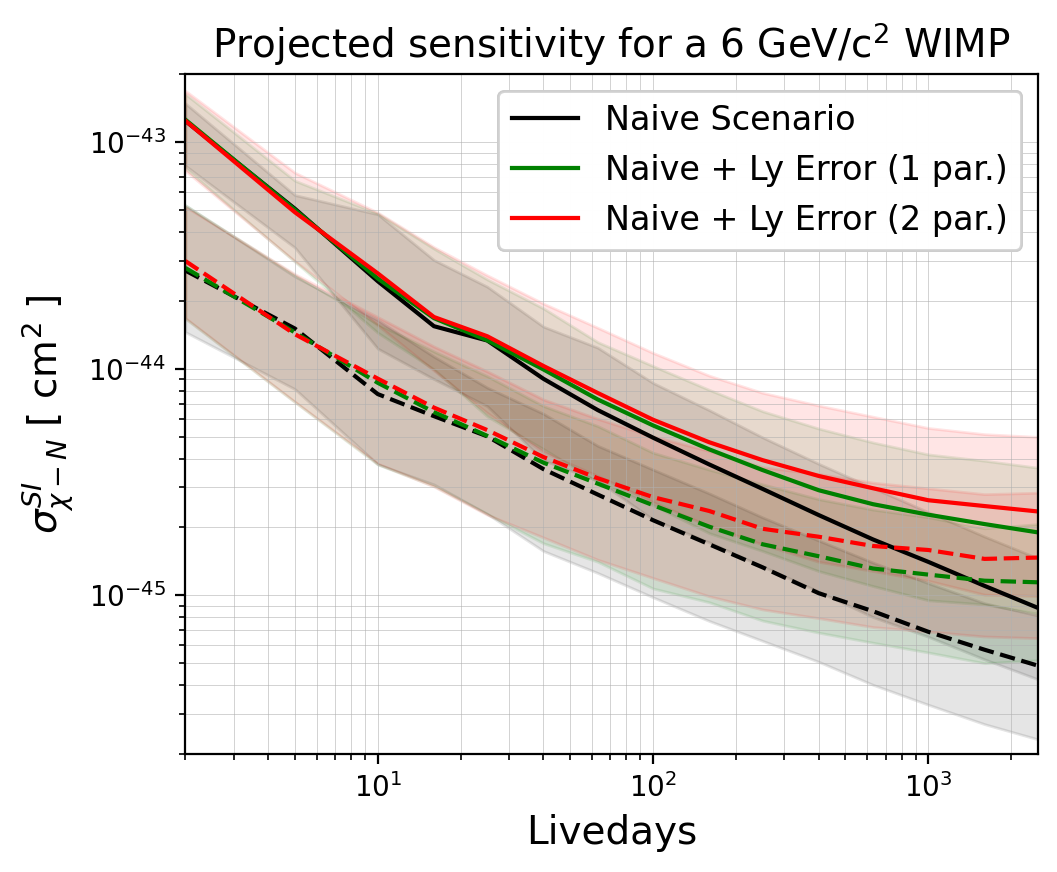}
        \caption{The projected 6 GeV/c$^2$ WIMP sensitivity (the median 90\% exclusion limit) for a 5.6-ton LZ-like detector as a function of live days. The solid curves are the detector threshold of 3-fold S1 coincidence and 5 \Nee S2 requirement, and the dashed curves are a relaxed 2-fold 5 \Nee detector threshold. The black curves are the naive scenario where yield uncertainty is not included. The shaded bands are the 68\% overall uncertainties from PLR calculation.}
        \label{fig:limit_vs_time}
    \end{center}
\end{figure}

The effect of yield uncertainties on projected WIMP sensitivities as a function of accumulated exposure is shown in Fig. \ref{fig:limit_vs_time}, where the color scheme is the same as used in Fig. \ref{fig:limit_vs_mass}. In the ideal scenario of zero yield uncertainties, the projected WIMP sensitivity continues to improve without any signs of saturation as the exposure increases.
However, when \ly uncertainties are introduced, the projected sensitivity of the hypothetical experiment to 6 GeV/c$^2$ WIMP begins to saturate around an exposure of \SI{30.7}{tonne \cdot year}. This effect is not as significant at low exposure values because the expected $^8$B rate is small and the sensitivity is mostly limited by Poisson fluctuation. The transition between the two regimes occurs at $\sim$\SI{700}{tonne \cdot day}. LZ is currently the leading LXe dark matter experiment with a reported exposure of \SI{330}{tonne \cdot day}~\cite{LZ:2022ufs}, and has not yet reached the saturation point. As LZ and other ton-scale LXe experiments continue to accumulate statistics, they may find it difficult to constrain the $^8$B background rate at $<$20\% accuracy without further reducing the \ly systematic uncertainty. 
Eventually this issue may cause them to approach the neutrino floor near the 6 GeV/c$^2$ mass sooner than previously expected.


\section{Discussion} \label{sec:discuss}

Traditionally LXe dark matter experiments mostly focus on the WIMP mass range of 10 GeV/$c^2$ to 1 TeV/$c^2$, but recent progresses have enabled them to start probing WIMP masses below 10 GeV/$c^2$~\cite{LUX:2019npm, LUX:2018akb, XENON:2019gfn}. In this mass region, a LXe experiment's WIMP sensitivity is negatively impacted by the presence of $^8$B \cns\ interactions, which is further amplified by the large uncertainties in measured NR yields as demonstrated in Sec.~\ref{sec:phys}. 
In addition, the yield uncertainties also jeopardize the capability of an experiment to accurately measure the $^8$B \cns\ interaction properties, which may 
potentially reveal new physics beyond the Standard Model predictions. Specifically, a deficit in $^8$B flux could be interpreted as evidence for active-to-sterile neutrino oscillation~\cite{Billard:2014yka}, while an excess may suggest non-standard neutrino interaction (NSI)~\cite{COHERENT:2020iec,Brice:2013fwa}. Hence, new NR calibrations that can significantly reduce uncertainties in the yields, especially for \ly, can significantly improve the physics reach of current and future LXe experiments. 

\begin{figure}[!!t]
    \begin{center}
        \includegraphics[width=0.5\textwidth]{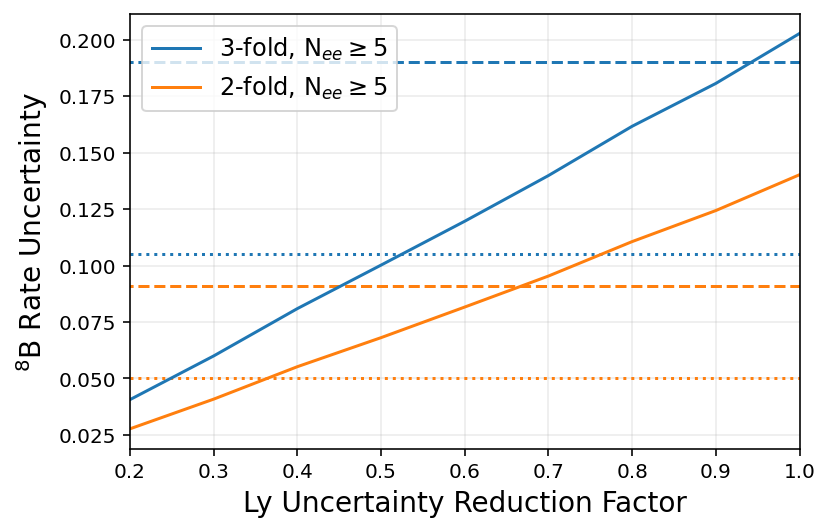}
        \caption{Calibration requirement of the \ly uncertainties for the $^8$B rate at two different thresholds. The dashed lines are the corresponding levels of Poisson uncertainties for $^8$B neutrino in a projected 15.3 \ty exposure. The dotted lines are for a 50 \ty exposure. }
        \label{fig:ly_calibration}
    \end{center}
\end{figure}

Figure~\ref{fig:ly_calibration} illustrates the estimated uncertainty of a $^8$B measurement assuming different levels of \ly calibration accuracy. In this study, we artificially reduce the overall \ly uncertainties reported in ~\cite{Huang:2020, LUX:2022qxb} by a chosen factor and then use the method explained in Sec. \ref{sec:phys:eff} to estimate the $^8$B rate uncertainty. 
Assuming a detector threshold of 3-fold S1 and 5-electron S2 and a projected 15.3 \ty full exposure for LZ, the \ly uncertainty needs to be reduced to 93\% of its current value in order to match the level of Poisson fluctuations, as illustrated by the horizontal dashed blue line. If a 2-fold S1 threshold is used, a stronger reduction factor of 0.67 will be required, as illustrated by the horizontal dashed orange line. Since the statistical uncertainty is proportional to the inverse square root of exposure, future LXe experiments would need even stricter calibration requirements to take full advantage of the larger exposures, as illustrated by the horizontal dotted lines in Fig.~\ref{fig:ly_calibration} for a hypothetical 50 \ty exposure experiment.

Elastic neutron scattering is a demonstrated method to produce nuclear recoils of desired energy distributions, and is the approach used in the most precise calibrations in LXe discussed in Sec.~\ref{sec:model:data}. 
To improve the \ly\ calibration accuracy in the keV energy region, low-energy mono-energetic neutrons, such as those produced by (p,n) interactions or through down-scattering of MeV DD fusion neutrons on a deuterium- or hydrogen-rich reflector~\cite{Verbus:2016sgw}, will be needed. 
To control the systematic uncertainties in such calibrations, neutron timing derived from the neutron beam or by instrumenting the down-scattering neutron reflector volumes can provide critical background rejection power. 
Photo-neutron sources such as YBe and SbBe have also been proposed for calibrations of dark matter detectors in the low-energy region. 
A well-designed calibration experiment with carefully-controlled systematics may provide the much-needed improvement in \ly to to recover the losses of light mass WIMP and $^8$B sensitivities in large LXe experiments. 

\section{Conclusion} \label{sec:summary}

We developed NEST-based models to parameterize the low-energy NR ionization and scintillation yields. We propagate the yield uncertainties into the WIMP sensitivity in the mass range of 4 to 10 GeV/$c^2$ in the presence of solar $^8$B neutrino \cns backgrounds. In this work, we primarily used calibration data from LUX Run3~\cite{LUX:2016ezw} and Run4~\cite{Huang:2020, LUX:2022qxb} to constrain the \ly uncertainty below 3.4 \kevnr, and data from Livermore~\cite{Lenardo:2019fcn} and LUX Run4~\cite{Huang:2020, LUX:2022qxb} to constrain the \qy uncertainty below 3.4 \kevnr. We found that the dominant systematical uncertainty in terms of the expected $^8$B \cns rate in LZ comes from the \ly, with a fractional uncertainty of $\sim20\%$ for 3-fold, 5 \Nee detector threshold, and $\sim14\%$ for 2-fold 5 \Nee threshold. The low-energy \qy systematic uncertainty in terms of predicted $^8$B rate is $\sim6\%$ for a $N_{ee}>5$ threshold and can be absorbed into the \ly nuisance parameter during the PLR calculation.

The yield uncertainties introduce correlations between $^8$B background and low-mass WIMP signals, which constructively amplify the impact on the WIMP sensitivity. In addition, we obtained a more conservative projections when we parameterized \ly with with the $(a,b)$ model (Sec. \ref{sec:model:ab}) than simply shifting the \ly up and down (Sec. \ref{sec:model:a}), suggesting more than one degree of freedom is needed to characterize the \ly uncertainty.

The impact of yield uncertainties on WIMP sensitivity is the strongest at 5.5-6.0 GeV/c$^2$, but diminishes rather quickly toward 10 GeV. The effect of yield uncertainties is subdominant in smaller exposures ($<1.5$ \ty) due to Poisson fluctuations. However, the long-term impact becomes more severe as the exposure increase, ultimately reaching saturation for 6 GeV/c$^2$ WIMP around 20 \ty. Hence, it is crucial for the next generation LXe dark matter experiment to carry out better low-energy NR calibrations, especially for constraining the uncertainty in \ly.

\begin{acknowledgments}

This work used two software packages developed by the LZ Collaboration --- \textsc{LZNESTUtils}, a NEST interfaces package, and \textsc{LZStats}, a profile likelihood ratio statistical package. This work used computing resources from the Center for Computation and Visualization provided by Brown University, and from the National Energy Research Scientific Computing Center, a DOE Office of Science User Facility supported by the Office of Science of the U.S. Department of Energy under Contract No. DE-AC02-05CH11231. The authors would like to thank many NEST and LZ collaboration members for insightful discussion and feedback.

\end{acknowledgments}


\bibliography{bibtex_inspirehep}

\begin{thebibliography}{39}%
\makeatletter
\providecommand \@ifxundefined [1]{%
 \@ifx{#1\undefined}
}%
\providecommand \@ifnum [1]{%
 \ifnum #1\expandafter \@firstoftwo
 \else \expandafter \@secondoftwo
 \fi
}%
\providecommand \@ifx [1]{%
 \ifx #1\expandafter \@firstoftwo
 \else \expandafter \@secondoftwo
 \fi
}%
\providecommand \natexlab [1]{#1}%
\providecommand \enquote  [1]{``#1''}%
\providecommand \bibnamefont  [1]{#1}%
\providecommand \bibfnamefont [1]{#1}%
\providecommand \citenamefont [1]{#1}%
\providecommand \href@noop [0]{\@secondoftwo}%
\providecommand \href [0]{\begingroup \@sanitize@url \@href}%
\providecommand \@href[1]{\@@startlink{#1}\@@href}%
\providecommand \@@href[1]{\endgroup#1\@@endlink}%
\providecommand \@sanitize@url [0]{\catcode `\\12\catcode `\$12\catcode
  `\&12\catcode `\#12\catcode `\^12\catcode `\_12\catcode `\%12\relax}%
\providecommand \@@startlink[1]{}%
\providecommand \@@endlink[0]{}%
\providecommand \url  [0]{\begingroup\@sanitize@url \@url }%
\providecommand \@url [1]{\endgroup\@href {#1}{\urlprefix }}%
\providecommand \urlprefix  [0]{URL }%
\providecommand \Eprint [0]{\href }%
\providecommand \doibase [0]{http://dx.doi.org/}%
\providecommand \selectlanguage [0]{\@gobble}%
\providecommand \bibinfo  [0]{\@secondoftwo}%
\providecommand \bibfield  [0]{\@secondoftwo}%
\providecommand \translation [1]{[#1]}%
\providecommand \BibitemOpen [0]{}%
\providecommand \bibitemStop [0]{}%
\providecommand \bibitemNoStop [0]{.\EOS\space}%
\providecommand \EOS [0]{\spacefactor3000\relax}%
\providecommand \BibitemShut  [1]{\csname bibitem#1\endcsname}%
\let\auto@bib@innerbib\@empty
\bibitem [{\citenamefont {Aalbers}\ \emph
  {et~al.}(2022{\natexlab{a}})\citenamefont {Aalbers} \emph
  {et~al.}}]{LZ:2022ufs}%
  \BibitemOpen
  \bibfield  {author} {\bibinfo {author} {\bibfnamefont {J.}~\bibnamefont
  {Aalbers}} \emph {et~al.} (\bibinfo {collaboration} {LZ}),\ }\href@noop {} {\
   (\bibinfo {year} {2022}{\natexlab{a}})},\ \Eprint
  {http://arxiv.org/abs/2207.03764} {arXiv:2207.03764 [hep-ex]} \BibitemShut
  {NoStop}%
\bibitem [{\citenamefont {Aprile}\ \emph {et~al.}(2020)\citenamefont {Aprile}
  \emph {et~al.}}]{XENON:2020kmp}%
  \BibitemOpen
  \bibfield  {author} {\bibinfo {author} {\bibfnamefont {E.}~\bibnamefont
  {Aprile}} \emph {et~al.} (\bibinfo {collaboration} {XENON}),\ }\href
  {\doibase 10.1088/1475-7516/2020/11/031} {\bibfield  {journal} {\bibinfo
  {journal} {JCAP}\ }\textbf {\bibinfo {volume} {11}},\ \bibinfo {pages} {031}
  (\bibinfo {year} {2020})},\ \Eprint {http://arxiv.org/abs/2007.08796}
  {arXiv:2007.08796 [physics.ins-det]} \BibitemShut {NoStop}%
\bibitem [{\citenamefont {Meng}\ \emph {et~al.}(2021)\citenamefont {Meng} \emph
  {et~al.}}]{PandaX-4T:2021bab}%
  \BibitemOpen
  \bibfield  {author} {\bibinfo {author} {\bibfnamefont {Y.}~\bibnamefont
  {Meng}} \emph {et~al.} (\bibinfo {collaboration} {PandaX-4T}),\ }\href
  {\doibase 10.1103/PhysRevLett.127.261802} {\bibfield  {journal} {\bibinfo
  {journal} {Phys. Rev. Lett.}\ }\textbf {\bibinfo {volume} {127}},\ \bibinfo
  {pages} {261802} (\bibinfo {year} {2021})},\ \Eprint
  {http://arxiv.org/abs/2107.13438} {arXiv:2107.13438 [hep-ex]} \BibitemShut
  {NoStop}%
\bibitem [{\citenamefont {Aalbers}\ \emph
  {et~al.}(2022{\natexlab{b}})\citenamefont {Aalbers} \emph
  {et~al.}}]{Aalbers:2022dzr}%
  \BibitemOpen
  \bibfield  {author} {\bibinfo {author} {\bibfnamefont {J.}~\bibnamefont
  {Aalbers}} \emph {et~al.},\ }\href@noop {} {\  (\bibinfo {year}
  {2022}{\natexlab{b}})},\ \Eprint {http://arxiv.org/abs/2203.02309}
  {arXiv:2203.02309 [physics.ins-det]} \BibitemShut {NoStop}%
\bibitem [{\citenamefont {Lenardo}\ \emph {et~al.}(2015)\citenamefont
  {Lenardo}, \citenamefont {Kazkaz}, \citenamefont {Manalaysay}, \citenamefont
  {Mock}, \citenamefont {Szydagis},\ and\ \citenamefont
  {Tripathi}}]{Lenardo:2015}%
  \BibitemOpen
  \bibfield  {author} {\bibinfo {author} {\bibfnamefont {B.}~\bibnamefont
  {Lenardo}}, \bibinfo {author} {\bibfnamefont {K.}~\bibnamefont {Kazkaz}},
  \bibinfo {author} {\bibfnamefont {A.}~\bibnamefont {Manalaysay}}, \bibinfo
  {author} {\bibfnamefont {J.}~\bibnamefont {Mock}}, \bibinfo {author}
  {\bibfnamefont {M.}~\bibnamefont {Szydagis}}, \ and\ \bibinfo {author}
  {\bibfnamefont {M.}~\bibnamefont {Tripathi}},\ }\href {\doibase
  10.1109/TNS.2015.2481322} {\bibfield  {journal} {\bibinfo  {journal} {IEEE
  Transactions on Nuclear Science}\ }\textbf {\bibinfo {volume} {62}},\
  \bibinfo {pages} {3387} (\bibinfo {year} {2015})}\BibitemShut {NoStop}%
\bibitem [{\citenamefont {Freedman}(1974)}]{Freedman:1973yd}%
  \BibitemOpen
  \bibfield  {author} {\bibinfo {author} {\bibfnamefont {D.~Z.}\ \bibnamefont
  {Freedman}},\ }\href {\doibase 10.1103/PhysRevD.9.1389} {\bibfield  {journal}
  {\bibinfo  {journal} {Phys. Rev. D}\ }\textbf {\bibinfo {volume} {9}},\
  \bibinfo {pages} {1389} (\bibinfo {year} {1974})}\BibitemShut {NoStop}%
\bibitem [{\citenamefont {Akimov}\ \emph {et~al.}(2017)\citenamefont {Akimov}
  \emph {et~al.}}]{COHERENT:2017ipa}%
  \BibitemOpen
  \bibfield  {author} {\bibinfo {author} {\bibfnamefont {D.}~\bibnamefont
  {Akimov}} \emph {et~al.} (\bibinfo {collaboration} {COHERENT}),\ }\href
  {\doibase 10.1126/science.aao0990} {\bibfield  {journal} {\bibinfo  {journal}
  {Science}\ }\textbf {\bibinfo {volume} {357}},\ \bibinfo {pages} {1123}
  (\bibinfo {year} {2017})},\ \Eprint {http://arxiv.org/abs/1708.01294}
  {arXiv:1708.01294 [nucl-ex]} \BibitemShut {NoStop}%
\bibitem [{\citenamefont {Aharmim}\ \emph {et~al.}(2013)\citenamefont {Aharmim}
  \emph {et~al.}}]{SNO:2011hxd}%
  \BibitemOpen
  \bibfield  {author} {\bibinfo {author} {\bibfnamefont {B.}~\bibnamefont
  {Aharmim}} \emph {et~al.} (\bibinfo {collaboration} {SNO}),\ }\href {\doibase
  10.1103/PhysRevC.88.025501} {\bibfield  {journal} {\bibinfo  {journal} {Phys.
  Rev. C}\ }\textbf {\bibinfo {volume} {88}},\ \bibinfo {pages} {025501}
  (\bibinfo {year} {2013})},\ \Eprint {http://arxiv.org/abs/1109.0763}
  {arXiv:1109.0763 [nucl-ex]} \BibitemShut {NoStop}%
\bibitem [{\citenamefont {Baxter}\ \emph {et~al.}(2021)\citenamefont {Baxter}
  \emph {et~al.}}]{Baxter:2021pqo}%
  \BibitemOpen
  \bibfield  {author} {\bibinfo {author} {\bibfnamefont {D.}~\bibnamefont
  {Baxter}} \emph {et~al.},\ }\href {\doibase 10.1140/epjc/s10052-021-09655-y}
  {\bibfield  {journal} {\bibinfo  {journal} {Eur. Phys. J. C}\ }\textbf
  {\bibinfo {volume} {81}},\ \bibinfo {pages} {907} (\bibinfo {year} {2021})},\
  \Eprint {http://arxiv.org/abs/2105.00599} {arXiv:2105.00599 [hep-ex]}
  \BibitemShut {NoStop}%
\bibitem [{\citenamefont {Strigari}(2009)}]{Strigari:2009bq}%
  \BibitemOpen
  \bibfield  {author} {\bibinfo {author} {\bibfnamefont {L.~E.}\ \bibnamefont
  {Strigari}},\ }\href {\doibase 10.1088/1367-2630/11/10/105011} {\bibfield
  {journal} {\bibinfo  {journal} {New J. Phys.}\ }\textbf {\bibinfo {volume}
  {11}},\ \bibinfo {pages} {105011} (\bibinfo {year} {2009})},\ \Eprint
  {http://arxiv.org/abs/0903.3630} {arXiv:0903.3630 [astro-ph.CO]} \BibitemShut
  {NoStop}%
\bibitem [{\citenamefont {Honda}\ \emph {et~al.}(2011)\citenamefont {Honda},
  \citenamefont {Kajita}, \citenamefont {Kasahara},\ and\ \citenamefont
  {Midorikawa}}]{Honda:2011nf}%
  \BibitemOpen
  \bibfield  {author} {\bibinfo {author} {\bibfnamefont {M.}~\bibnamefont
  {Honda}}, \bibinfo {author} {\bibfnamefont {T.}~\bibnamefont {Kajita}},
  \bibinfo {author} {\bibfnamefont {K.}~\bibnamefont {Kasahara}}, \ and\
  \bibinfo {author} {\bibfnamefont {S.}~\bibnamefont {Midorikawa}},\ }\href
  {\doibase 10.1103/PhysRevD.83.123001} {\bibfield  {journal} {\bibinfo
  {journal} {Phys. Rev. D}\ }\textbf {\bibinfo {volume} {83}},\ \bibinfo
  {pages} {123001} (\bibinfo {year} {2011})},\ \Eprint
  {http://arxiv.org/abs/1102.2688} {arXiv:1102.2688 [astro-ph.HE]} \BibitemShut
  {NoStop}%
\bibitem [{\citenamefont {Richard}\ \emph {et~al.}(2016)\citenamefont {Richard}
  \emph {et~al.}}]{Super-Kamiokande:2015qek}%
  \BibitemOpen
  \bibfield  {author} {\bibinfo {author} {\bibfnamefont {E.}~\bibnamefont
  {Richard}} \emph {et~al.} (\bibinfo {collaboration} {Super-Kamiokande}),\
  }\href {\doibase 10.1103/PhysRevD.94.052001} {\bibfield  {journal} {\bibinfo
  {journal} {Phys. Rev. D}\ }\textbf {\bibinfo {volume} {94}},\ \bibinfo
  {pages} {052001} (\bibinfo {year} {2016})},\ \Eprint
  {http://arxiv.org/abs/1510.08127} {arXiv:1510.08127 [hep-ex]} \BibitemShut
  {NoStop}%
\bibitem [{\citenamefont {Aprile}\ \emph {et~al.}(2021)\citenamefont {Aprile}
  \emph {et~al.}}]{XENON:2020gfr}%
  \BibitemOpen
  \bibfield  {author} {\bibinfo {author} {\bibfnamefont {E.}~\bibnamefont
  {Aprile}} \emph {et~al.} (\bibinfo {collaboration} {XENON}),\ }\href
  {\doibase 10.1103/PhysRevLett.126.091301} {\bibfield  {journal} {\bibinfo
  {journal} {Phys. Rev. Lett.}\ }\textbf {\bibinfo {volume} {126}},\ \bibinfo
  {pages} {091301} (\bibinfo {year} {2021})},\ \Eprint
  {http://arxiv.org/abs/2012.02846} {arXiv:2012.02846 [hep-ex]} \BibitemShut
  {NoStop}%
\bibitem [{\citenamefont {Akerib}\ \emph
  {et~al.}(2020{\natexlab{a}})\citenamefont {Akerib} \emph
  {et~al.}}]{LUX-ZEPLIN:2018poe}%
  \BibitemOpen
  \bibfield  {author} {\bibinfo {author} {\bibfnamefont {D.~S.}\ \bibnamefont
  {Akerib}} \emph {et~al.} (\bibinfo {collaboration} {LUX-ZEPLIN}),\ }\href
  {\doibase 10.1103/PhysRevD.101.052002} {\bibfield  {journal} {\bibinfo
  {journal} {Phys. Rev. D}\ }\textbf {\bibinfo {volume} {101}},\ \bibinfo
  {pages} {052002} (\bibinfo {year} {2020}{\natexlab{a}})},\ \Eprint
  {http://arxiv.org/abs/1802.06039} {arXiv:1802.06039 [astro-ph.IM]}
  \BibitemShut {NoStop}%
\bibitem [{\citenamefont {Akimov}\ \emph {et~al.}(2021)\citenamefont {Akimov}
  \emph {et~al.}}]{COHERENT:2020iec}%
  \BibitemOpen
  \bibfield  {author} {\bibinfo {author} {\bibfnamefont {D.}~\bibnamefont
  {Akimov}} \emph {et~al.} (\bibinfo {collaboration} {COHERENT}),\ }\href
  {\doibase 10.1103/PhysRevLett.126.012002} {\bibfield  {journal} {\bibinfo
  {journal} {Phys. Rev. Lett.}\ }\textbf {\bibinfo {volume} {126}},\ \bibinfo
  {pages} {012002} (\bibinfo {year} {2021})},\ \Eprint
  {http://arxiv.org/abs/2003.10630} {arXiv:2003.10630 [nucl-ex]} \BibitemShut
  {NoStop}%
\bibitem [{\citenamefont {Brice}\ \emph {et~al.}(2014)\citenamefont {Brice}
  \emph {et~al.}}]{Brice:2013fwa}%
  \BibitemOpen
  \bibfield  {author} {\bibinfo {author} {\bibfnamefont {S.~J.}\ \bibnamefont
  {Brice}} \emph {et~al.},\ }\href {\doibase 10.1103/PhysRevD.89.072004}
  {\bibfield  {journal} {\bibinfo  {journal} {Phys. Rev. D}\ }\textbf {\bibinfo
  {volume} {89}},\ \bibinfo {pages} {072004} (\bibinfo {year} {2014})},\
  \Eprint {http://arxiv.org/abs/1311.5958} {arXiv:1311.5958 [physics.ins-det]}
  \BibitemShut {NoStop}%
\bibitem [{\citenamefont {Freese}\ \emph {et~al.}(2013)\citenamefont {Freese},
  \citenamefont {Lisanti},\ and\ \citenamefont {Savage}}]{Freese:2012xd}%
  \BibitemOpen
  \bibfield  {author} {\bibinfo {author} {\bibfnamefont {K.}~\bibnamefont
  {Freese}}, \bibinfo {author} {\bibfnamefont {M.}~\bibnamefont {Lisanti}}, \
  and\ \bibinfo {author} {\bibfnamefont {C.}~\bibnamefont {Savage}},\ }\href
  {\doibase 10.1103/RevModPhys.85.1561} {\bibfield  {journal} {\bibinfo
  {journal} {Rev. Mod. Phys.}\ }\textbf {\bibinfo {volume} {85}},\ \bibinfo
  {pages} {1561} (\bibinfo {year} {2013})},\ \Eprint
  {http://arxiv.org/abs/1209.3339} {arXiv:1209.3339 [astro-ph.CO]} \BibitemShut
  {NoStop}%
\bibitem [{\citenamefont {Lindhard}\ \emph {et~al.}(1963)\citenamefont
  {Lindhard}, \citenamefont {Scharff},\ and\ \citenamefont
  {Schioett}}]{osti_4153115}%
  \BibitemOpen
  \bibfield  {author} {\bibinfo {author} {\bibfnamefont {J.}~\bibnamefont
  {Lindhard}}, \bibinfo {author} {\bibfnamefont {M.}~\bibnamefont {Scharff}}, \
  and\ \bibinfo {author} {\bibfnamefont {H.~E.}\ \bibnamefont {Schioett}},\
  }\href {https://www.osti.gov/biblio/4153115} {\bibfield  {journal} {\bibinfo
  {journal} {Kgl. Danske Videnskab. Selskab. Mat. Fys. Medd.}\ }\textbf
  {\bibinfo {volume} {Vol: 33: No. 14}} (\bibinfo {year} {1963})}\BibitemShut
  {NoStop}%
\bibitem [{\citenamefont {Szydagis}\ \emph {et~al.}(2021)\citenamefont
  {Szydagis}, \citenamefont {Block}, \citenamefont {Farquhar}, \citenamefont
  {Flesher}, \citenamefont {Kozlova}, \citenamefont {Levy}, \citenamefont
  {Mangus}, \citenamefont {Mooney}, \citenamefont {Mueller}, \citenamefont
  {Rischbieter},\ and\ \citenamefont {Schwartz}}]{Szydagis:2021}%
  \BibitemOpen
  \bibfield  {author} {\bibinfo {author} {\bibfnamefont {M.}~\bibnamefont
  {Szydagis}}, \bibinfo {author} {\bibfnamefont {G.~A.}\ \bibnamefont {Block}},
  \bibinfo {author} {\bibfnamefont {C.}~\bibnamefont {Farquhar}}, \bibinfo
  {author} {\bibfnamefont {A.~J.}\ \bibnamefont {Flesher}}, \bibinfo {author}
  {\bibfnamefont {E.~S.}\ \bibnamefont {Kozlova}}, \bibinfo {author}
  {\bibfnamefont {C.}~\bibnamefont {Levy}}, \bibinfo {author} {\bibfnamefont
  {E.~A.}\ \bibnamefont {Mangus}}, \bibinfo {author} {\bibfnamefont
  {M.}~\bibnamefont {Mooney}}, \bibinfo {author} {\bibfnamefont
  {J.}~\bibnamefont {Mueller}}, \bibinfo {author} {\bibfnamefont {G.~R.~C.}\
  \bibnamefont {Rischbieter}}, \ and\ \bibinfo {author} {\bibfnamefont {A.~K.}\
  \bibnamefont {Schwartz}},\ }\href {\doibase 10.3390/instruments5010013}
  {\bibfield  {journal} {\bibinfo  {journal} {Instruments}\ }\textbf {\bibinfo
  {volume} {5}} (\bibinfo {year} {2021}),\
  10.3390/instruments5010013}\BibitemShut {NoStop}%
\bibitem [{\citenamefont {Aprile}\ \emph
  {et~al.}(2018{\natexlab{a}})\citenamefont {Aprile}, \citenamefont {Anthony},
  \citenamefont {Lin}, \citenamefont {Greene}, \citenamefont {De~Perio},
  \citenamefont {Gao}, \citenamefont {Howlett}, \citenamefont {Plante},
  \citenamefont {Zhang},\ and\ \citenamefont {Zhu}}]{Aprile:2018jvg}%
  \BibitemOpen
  \bibfield  {author} {\bibinfo {author} {\bibfnamefont {E.}~\bibnamefont
  {Aprile}}, \bibinfo {author} {\bibfnamefont {M.}~\bibnamefont {Anthony}},
  \bibinfo {author} {\bibfnamefont {Q.}~\bibnamefont {Lin}}, \bibinfo {author}
  {\bibfnamefont {Z.}~\bibnamefont {Greene}}, \bibinfo {author} {\bibfnamefont
  {P.}~\bibnamefont {De~Perio}}, \bibinfo {author} {\bibfnamefont
  {F.}~\bibnamefont {Gao}}, \bibinfo {author} {\bibfnamefont {J.}~\bibnamefont
  {Howlett}}, \bibinfo {author} {\bibfnamefont {G.}~\bibnamefont {Plante}},
  \bibinfo {author} {\bibfnamefont {Y.}~\bibnamefont {Zhang}}, \ and\ \bibinfo
  {author} {\bibfnamefont {T.}~\bibnamefont {Zhu}},\ }\href {\doibase
  10.1103/PhysRevD.98.112003} {\bibfield  {journal} {\bibinfo  {journal} {Phys.
  Rev. D}\ }\textbf {\bibinfo {volume} {98}},\ \bibinfo {pages} {112003}
  (\bibinfo {year} {2018}{\natexlab{a}})},\ \Eprint
  {http://arxiv.org/abs/1809.02072} {arXiv:1809.02072 [physics.ins-det]}
  \BibitemShut {NoStop}%
\bibitem [{\citenamefont {Huang}(2020)}]{Huang:2020}%
  \BibitemOpen
  \bibfield  {author} {\bibinfo {author} {\bibfnamefont {D.}~\bibnamefont
  {Huang}},\ }\emph {\bibinfo {title} {Ultra-Low Energy Calibration of the LUX
  and LZ Dark Matter Detectors}},\ \href@noop {} {Ph.D. thesis},\ \bibinfo
  {school} {Brown University}, \bibinfo {address}
  {https://doi.org/10.26300/zvs6-fx07} (\bibinfo {year} {2020}),\ \bibinfo
  {note} {physics Theses and Dissertations.}\BibitemShut {Stop}%
\bibitem [{\citenamefont {Akerib}\ \emph {et~al.}(2022)\citenamefont {Akerib}
  \emph {et~al.}}]{LUX:2022qxb}%
  \BibitemOpen
  \bibfield  {author} {\bibinfo {author} {\bibfnamefont {D.~S.}\ \bibnamefont
  {Akerib}} \emph {et~al.} (\bibinfo {collaboration} {LUX}),\ }\href@noop {} {\
   (\bibinfo {year} {2022})},\ \Eprint {http://arxiv.org/abs/2210.05859}
  {arXiv:2210.05859 [physics.ins-det]} \BibitemShut {NoStop}%
\bibitem [{\citenamefont {Akerib}\ \emph {et~al.}(2016)\citenamefont {Akerib}
  \emph {et~al.}}]{LUX:2016ezw}%
  \BibitemOpen
  \bibfield  {author} {\bibinfo {author} {\bibfnamefont {D.~S.}\ \bibnamefont
  {Akerib}} \emph {et~al.} (\bibinfo {collaboration} {LUX}),\ }\href@noop {} {\
   (\bibinfo {year} {2016})},\ \Eprint {http://arxiv.org/abs/1608.05381}
  {arXiv:1608.05381 [physics.ins-det]} \BibitemShut {NoStop}%
\bibitem [{\citenamefont {Lenardo}\ \emph {et~al.}(2019)\citenamefont
  {Lenardo}, \citenamefont {Xu}, \citenamefont {Pereverzev}, \citenamefont
  {Akindele}, \citenamefont {Naim}, \citenamefont {Kingston}, \citenamefont
  {Bernstein}, \citenamefont {Kazkaz}, \citenamefont {Tripathi}, \citenamefont
  {Awe}, \citenamefont {Li}, \citenamefont {Runge}, \citenamefont {Hedges},
  \citenamefont {An},\ and\ \citenamefont {Barbeau}}]{Lenardo:2019fcn}%
  \BibitemOpen
  \bibfield  {author} {\bibinfo {author} {\bibfnamefont {B.~G.}\ \bibnamefont
  {Lenardo}}, \bibinfo {author} {\bibfnamefont {J.}~\bibnamefont {Xu}},
  \bibinfo {author} {\bibfnamefont {S.}~\bibnamefont {Pereverzev}}, \bibinfo
  {author} {\bibfnamefont {O.~A.}\ \bibnamefont {Akindele}}, \bibinfo {author}
  {\bibfnamefont {D.}~\bibnamefont {Naim}}, \bibinfo {author} {\bibfnamefont
  {J.}~\bibnamefont {Kingston}}, \bibinfo {author} {\bibfnamefont
  {A.}~\bibnamefont {Bernstein}}, \bibinfo {author} {\bibfnamefont
  {K.}~\bibnamefont {Kazkaz}}, \bibinfo {author} {\bibfnamefont
  {M.}~\bibnamefont {Tripathi}}, \bibinfo {author} {\bibfnamefont
  {C.}~\bibnamefont {Awe}}, \bibinfo {author} {\bibfnamefont {L.}~\bibnamefont
  {Li}}, \bibinfo {author} {\bibfnamefont {J.}~\bibnamefont {Runge}}, \bibinfo
  {author} {\bibfnamefont {S.}~\bibnamefont {Hedges}}, \bibinfo {author}
  {\bibfnamefont {P.}~\bibnamefont {An}}, \ and\ \bibinfo {author}
  {\bibfnamefont {P.~S.}\ \bibnamefont {Barbeau}},\ }\href {\doibase
  10.1103/PhysRevLett.123.231106} {\bibfield  {journal} {\bibinfo  {journal}
  {Phys. Rev. Lett.}\ }\textbf {\bibinfo {volume} {123}},\ \bibinfo {pages}
  {231106} (\bibinfo {year} {2019})}\BibitemShut {NoStop}%
\bibitem [{\citenamefont {James}\ and\ \citenamefont
  {Roos}(1975)}]{James:1975dr}%
  \BibitemOpen
  \bibfield  {author} {\bibinfo {author} {\bibfnamefont {F.}~\bibnamefont
  {James}}\ and\ \bibinfo {author} {\bibfnamefont {M.}~\bibnamefont {Roos}},\
  }\href {\doibase 10.1016/0010-4655(75)90039-9} {\bibfield  {journal}
  {\bibinfo  {journal} {Comput. Phys. Commun.}\ }\textbf {\bibinfo {volume}
  {10}},\ \bibinfo {pages} {343} (\bibinfo {year} {1975})}\BibitemShut
  {NoStop}%
\bibitem [{\citenamefont {Akerib}\ \emph
  {et~al.}(2021{\natexlab{a}})\citenamefont {Akerib} \emph
  {et~al.}}]{LUX:2020yym}%
  \BibitemOpen
  \bibfield  {author} {\bibinfo {author} {\bibfnamefont {D.~S.}\ \bibnamefont
  {Akerib}} \emph {et~al.} (\bibinfo {collaboration} {LUX}),\ }\href {\doibase
  10.1103/PhysRevD.104.012011} {\bibfield  {journal} {\bibinfo  {journal}
  {Phys. Rev. D}\ }\textbf {\bibinfo {volume} {104}},\ \bibinfo {pages}
  {012011} (\bibinfo {year} {2021}{\natexlab{a}})},\ \Eprint
  {http://arxiv.org/abs/2011.09602} {arXiv:2011.09602 [hep-ex]} \BibitemShut
  {NoStop}%
\bibitem [{\citenamefont {Akerib}\ \emph
  {et~al.}(2021{\natexlab{b}})\citenamefont {Akerib} \emph
  {et~al.}}]{Akerib:2021pfd}%
  \BibitemOpen
  \bibfield  {author} {\bibinfo {author} {\bibfnamefont {D.~S.}\ \bibnamefont
  {Akerib}} \emph {et~al.},\ }\href@noop {} {\  (\bibinfo {year}
  {2021}{\natexlab{b}})},\ \Eprint {http://arxiv.org/abs/2101.08753}
  {arXiv:2101.08753 [astro-ph.IM]} \BibitemShut {NoStop}%
\bibitem [{\citenamefont {Aprile}\ \emph
  {et~al.}(2018{\natexlab{b}})\citenamefont {Aprile} \emph
  {et~al.}}]{XENON:2018voc}%
  \BibitemOpen
  \bibfield  {author} {\bibinfo {author} {\bibfnamefont {E.}~\bibnamefont
  {Aprile}} \emph {et~al.} (\bibinfo {collaboration} {XENON}),\ }\href
  {\doibase 10.1103/PhysRevLett.121.111302} {\bibfield  {journal} {\bibinfo
  {journal} {Phys. Rev. Lett.}\ }\textbf {\bibinfo {volume} {121}},\ \bibinfo
  {pages} {111302} (\bibinfo {year} {2018}{\natexlab{b}})},\ \Eprint
  {http://arxiv.org/abs/1805.12562} {arXiv:1805.12562 [astro-ph.CO]}
  \BibitemShut {NoStop}%
\bibitem [{\citenamefont {Anderson}\ \emph {et~al.}(2019)\citenamefont
  {Anderson} \emph {et~al.}}]{SNO:2018fch}%
  \BibitemOpen
  \bibfield  {author} {\bibinfo {author} {\bibfnamefont {M.}~\bibnamefont
  {Anderson}} \emph {et~al.} (\bibinfo {collaboration} {SNO+}),\ }\href
  {\doibase 10.1103/PhysRevD.99.012012} {\bibfield  {journal} {\bibinfo
  {journal} {Phys. Rev. D}\ }\textbf {\bibinfo {volume} {99}},\ \bibinfo
  {pages} {012012} (\bibinfo {year} {2019})},\ \Eprint
  {http://arxiv.org/abs/1812.03355} {arXiv:1812.03355 [hep-ex]} \BibitemShut
  {NoStop}%
\bibitem [{\citenamefont {Lewin}\ and\ \citenamefont
  {Smith}(1996)}]{Lewin:1995rx}%
  \BibitemOpen
  \bibfield  {author} {\bibinfo {author} {\bibfnamefont {J.~D.}\ \bibnamefont
  {Lewin}}\ and\ \bibinfo {author} {\bibfnamefont {P.~F.}\ \bibnamefont
  {Smith}},\ }\href {\doibase 10.1016/S0927-6505(96)00047-3} {\bibfield
  {journal} {\bibinfo  {journal} {Astropart. Phys.}\ }\textbf {\bibinfo
  {volume} {6}},\ \bibinfo {pages} {87} (\bibinfo {year} {1996})}\BibitemShut
  {NoStop}%
\bibitem [{\citenamefont {McCabe}(2014)}]{McCabe:2013kea}%
  \BibitemOpen
  \bibfield  {author} {\bibinfo {author} {\bibfnamefont {C.}~\bibnamefont
  {McCabe}},\ }\href {\doibase 10.1088/1475-7516/2014/02/027} {\bibfield
  {journal} {\bibinfo  {journal} {JCAP}\ }\textbf {\bibinfo {volume} {02}},\
  \bibinfo {pages} {027} (\bibinfo {year} {2014})},\ \Eprint
  {http://arxiv.org/abs/1312.1355} {arXiv:1312.1355 [astro-ph.CO]} \BibitemShut
  {NoStop}%
\bibitem [{\citenamefont {Smith}\ \emph {et~al.}(2007)\citenamefont {Smith}
  \emph {et~al.}}]{Smith:2006ym}%
  \BibitemOpen
  \bibfield  {author} {\bibinfo {author} {\bibfnamefont {M.~C.}\ \bibnamefont
  {Smith}} \emph {et~al.},\ }\href {\doibase 10.1111/j.1365-2966.2007.11964.x}
  {\bibfield  {journal} {\bibinfo  {journal} {Mon. Not. Roy. Astron. Soc.}\
  }\textbf {\bibinfo {volume} {379}},\ \bibinfo {pages} {755} (\bibinfo {year}
  {2007})},\ \Eprint {http://arxiv.org/abs/astro-ph/0611671}
  {arXiv:astro-ph/0611671} \BibitemShut {NoStop}%
\bibitem [{\citenamefont {Schoenrich}\ \emph {et~al.}(2010)\citenamefont
  {Schoenrich}, \citenamefont {Binney},\ and\ \citenamefont
  {Dehnen}}]{Schoenrich:2009bx}%
  \BibitemOpen
  \bibfield  {author} {\bibinfo {author} {\bibfnamefont {R.}~\bibnamefont
  {Schoenrich}}, \bibinfo {author} {\bibfnamefont {J.}~\bibnamefont {Binney}},
  \ and\ \bibinfo {author} {\bibfnamefont {W.}~\bibnamefont {Dehnen}},\ }\href
  {\doibase 10.1111/j.1365-2966.2010.16253.x} {\bibfield  {journal} {\bibinfo
  {journal} {Mon. Not. Roy. Astron. Soc.}\ }\textbf {\bibinfo {volume} {403}},\
  \bibinfo {pages} {1829} (\bibinfo {year} {2010})},\ \Eprint
  {http://arxiv.org/abs/0912.3693} {arXiv:0912.3693 [astro-ph.GA]} \BibitemShut
  {NoStop}%
\bibitem [{\citenamefont {Bland-Hawthorn}\ and\ \citenamefont
  {Gerhard}(2016)}]{Bland_Hawthorn_2016}%
  \BibitemOpen
  \bibfield  {author} {\bibinfo {author} {\bibfnamefont {J.}~\bibnamefont
  {Bland-Hawthorn}}\ and\ \bibinfo {author} {\bibfnamefont {O.}~\bibnamefont
  {Gerhard}},\ }\href {\doibase 10.1146/annurev-astro-081915-023441} {\bibfield
   {journal} {\bibinfo  {journal} {Annual Review of Astronomy and
  Astrophysics}\ }\textbf {\bibinfo {volume} {54}},\ \bibinfo {pages} {529}
  (\bibinfo {year} {2016})}\BibitemShut {NoStop}%
\bibitem [{\citenamefont {Akerib}\ \emph
  {et~al.}(2020{\natexlab{b}})\citenamefont {Akerib} \emph
  {et~al.}}]{LUX:2019npm}%
  \BibitemOpen
  \bibfield  {author} {\bibinfo {author} {\bibfnamefont {D.~S.}\ \bibnamefont
  {Akerib}} \emph {et~al.} (\bibinfo {collaboration} {LUX}),\ }\href {\doibase
  10.1103/PhysRevD.101.042001} {\bibfield  {journal} {\bibinfo  {journal}
  {Phys. Rev. D}\ }\textbf {\bibinfo {volume} {101}},\ \bibinfo {pages}
  {042001} (\bibinfo {year} {2020}{\natexlab{b}})},\ \Eprint
  {http://arxiv.org/abs/1907.06272} {arXiv:1907.06272 [astro-ph.CO]}
  \BibitemShut {NoStop}%
\bibitem [{\citenamefont {Akerib}\ \emph {et~al.}(2019)\citenamefont {Akerib}
  \emph {et~al.}}]{LUX:2018akb}%
  \BibitemOpen
  \bibfield  {author} {\bibinfo {author} {\bibfnamefont {D.~S.}\ \bibnamefont
  {Akerib}} \emph {et~al.} (\bibinfo {collaboration} {LUX}),\ }\href {\doibase
  10.1103/PhysRevLett.122.131301} {\bibfield  {journal} {\bibinfo  {journal}
  {Phys. Rev. Lett.}\ }\textbf {\bibinfo {volume} {122}},\ \bibinfo {pages}
  {131301} (\bibinfo {year} {2019})},\ \Eprint
  {http://arxiv.org/abs/1811.11241} {arXiv:1811.11241 [astro-ph.CO]}
  \BibitemShut {NoStop}%
\bibitem [{\citenamefont {Aprile}\ \emph {et~al.}(2019)\citenamefont {Aprile}
  \emph {et~al.}}]{XENON:2019gfn}%
  \BibitemOpen
  \bibfield  {author} {\bibinfo {author} {\bibfnamefont {E.}~\bibnamefont
  {Aprile}} \emph {et~al.} (\bibinfo {collaboration} {XENON}),\ }\href
  {\doibase 10.1103/PhysRevLett.123.251801} {\bibfield  {journal} {\bibinfo
  {journal} {Phys. Rev. Lett.}\ }\textbf {\bibinfo {volume} {123}},\ \bibinfo
  {pages} {251801} (\bibinfo {year} {2019})},\ \Eprint
  {http://arxiv.org/abs/1907.11485} {arXiv:1907.11485 [hep-ex]} \BibitemShut
  {NoStop}%
\bibitem [{\citenamefont {Billard}\ \emph {et~al.}(2015)\citenamefont
  {Billard}, \citenamefont {Strigari},\ and\ \citenamefont
  {Figueroa-Feliciano}}]{Billard:2014yka}%
  \BibitemOpen
  \bibfield  {author} {\bibinfo {author} {\bibfnamefont {J.}~\bibnamefont
  {Billard}}, \bibinfo {author} {\bibfnamefont {L.~E.}\ \bibnamefont
  {Strigari}}, \ and\ \bibinfo {author} {\bibfnamefont {E.}~\bibnamefont
  {Figueroa-Feliciano}},\ }\href {\doibase 10.1103/PhysRevD.91.095023}
  {\bibfield  {journal} {\bibinfo  {journal} {Phys. Rev. D}\ }\textbf {\bibinfo
  {volume} {91}},\ \bibinfo {pages} {095023} (\bibinfo {year} {2015})},\
  \Eprint {http://arxiv.org/abs/1409.0050} {arXiv:1409.0050 [astro-ph.CO]}
  \BibitemShut {NoStop}%
\bibitem [{\citenamefont {Verbus}\ \emph {et~al.}(2017)\citenamefont {Verbus}
  \emph {et~al.}}]{Verbus:2016sgw}%
  \BibitemOpen
  \bibfield  {author} {\bibinfo {author} {\bibfnamefont {J.~R.}\ \bibnamefont
  {Verbus}} \emph {et~al.},\ }\href {\doibase 10.1016/j.nima.2017.01.053}
  {\bibfield  {journal} {\bibinfo  {journal} {Nucl. Instrum. Meth. A}\ }\textbf
  {\bibinfo {volume} {851}},\ \bibinfo {pages} {68} (\bibinfo {year} {2017})},\
  \Eprint {http://arxiv.org/abs/1608.05309} {arXiv:1608.05309
  [physics.ins-det]} \BibitemShut {NoStop}%
\end{thebibliography}%

\appendix

\section{Simplified yield model uncertainties}

\begin{figure}[!!t]
    \begin{center}
        \includegraphics[width=.45\textwidth]{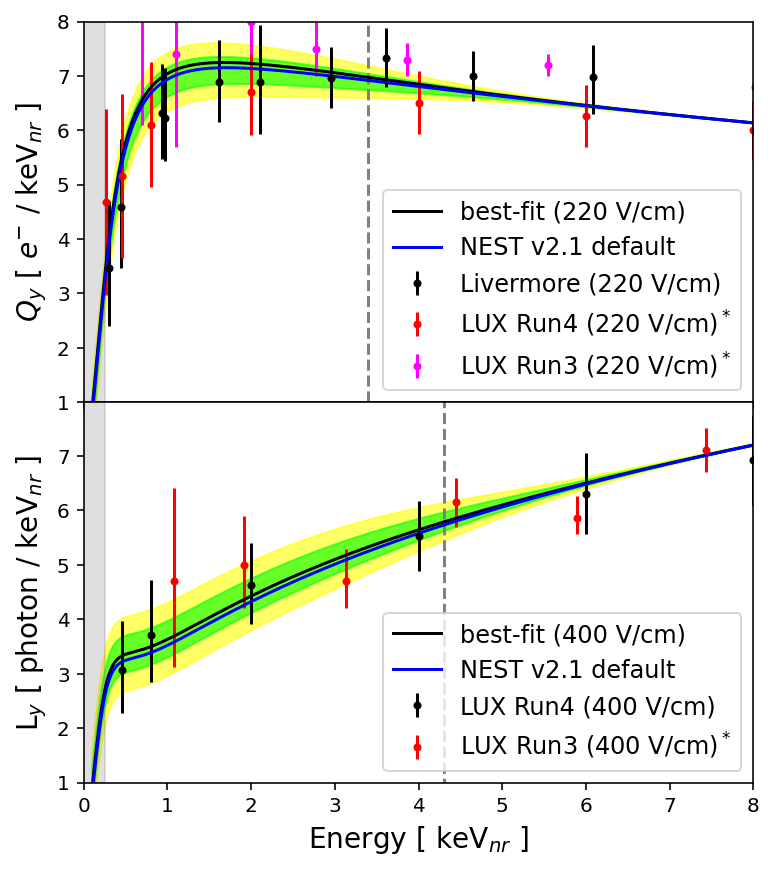}
        \caption{The Brazilian flag 1 and 2 sigma error bands for \qy and \ly propagated from the single parameter model model, similar to Fig. \ref{fig:lyqy_brazilian_flag_1par}}
        \label{fig:lyqy_brazilian_flag_1par}
    \end{center}
\end{figure}

One can derive the error envelope by varying one parameter per yield. For instance, Sec. \ref{sec:model:a} illustrate an alternative method by shifting the yield curves up and down by a constant. This method generates a relatively conservative error bands as shown in Fig. \ref{fig:lyqy_brazilian_flag_1par} in comparison to the $(a,b)$ parameterization. The fitting procedure and error propagation are identical to the description in the main paragraph. There are implicit shape variations that are not characterized by simply having one parameter shifting the NEST yield curve up and down. However, the shape variation only becomes significant when a LXe experiment is searching physics at very large exposures. It's convenient for the current generation LXe experiments to stick with one parameter variation, since they are likely to report low-energy physics results at a lower exposure. Unless \ly calibration is significantly improved, it's recommended for the next generation experiment to incorporate two parameters in characterizing \ly uncertainties when calculating the physics results. Furthermore, the one parameter method is likely sufficient for an LXe experiment that is overwhelmed by other sources of uncertainties in the same low-energy region, such as the instrumental background rising from random coincidence of phony S1s and spurious S2s.

\end{document}